\documentclass[fp,twocolumn]{jpsj3}
\usepackage{txfonts}

\usepackage{braket,bm}
\usepackage{graphicx}
\usepackage{color}

\usepackage{ulem}

\newcommand{\CenterRow}[2]{
  \dimen0=\ht\strutbox%
  \advance\dimen0\dp\strutbox%
  \multiply\dimen0 by#1%
  \divide\dimen0 by2%
  \advance\dimen0 by-.5\normalbaselineskip%
  \raisebox{-\dimen0}[0pt][0pt]{#2}}

\usepackage{ulem}
\usepackage{xcolor}
\DeclareRobustCommand{\erase}{\bgroup\markoverwith{\textcolor{red}{\rule[.5ex]{2pt}{0.4pt}}}\ULon}

\usepackage{here}

\title{Multifractality and Hyperuniformity in Quasicrystalline Bose-Hubbard Models with and without Disorder}

\author{Masahiro Hori$^{1,2}$\thanks{mhorijapan@gmail.com}, Takanori Sugimoto$^3$, Yoichiro Hashizume$^4$, and Takami Tohyama$^2$}
\inst{
$^1$Department of Physics and Engineering Physics, and Centre for Quantum Topology and Its Applications (quanTA), University of Saskatchewan, 116 Science Place, Saskatoon, Saskatchewan, Canada S7N 5E2 \\
$^2$Department of Applied Physics, Tokyo University of Science, Katsushika, Tokyo 125-8585, Japan \\
$^3$Center for Quantum Information and Quantum Biology, Osaka University, Osaka 560-0043, Japan \\
$^4$Institute of Arts and Sciences, Oshamambe Division, Tokyo University of Science, Hokkaido 049-3514, Japan
} 

\abst{
Clarifying similarities and differences in physical properties between crystalline and quasicrystalline systems is one of central issues in studying quasicrystals.
To contribute to this, we apply multifractal and hyperuniform analyses to nonuniform spatial patterns in the Bose-Hubbard model on the Penrose and Ammann-Beenker tilings.
Based on the mean-field approximation, we obtain real-space distributions of local superfluid amplitude and boson density.
In both Mott insulating and superfluid phases, the distributions are hyperuniform.
Analyzing the order metric that quantifies the complexity of nonuniform spatial patterns, we find that both quasicrystals show a significant increase of the order metric at a phase boundary between the Mott insulating and superfluid phases, in stark contrast to the case of a periodic square lattice.
Our results suggest that hyperuniformity is a useful concept to differentiate between crystalline and quasicrystalline bosonic systems.
The order metric clarifies if the distribution of a physical quantity reflects the point distribution or not, and quantifies how complex the distribution is in comparison with the point distribution.
Moreover, we introduce on-site random potentials into these quasicrystalline Bose-Hubbard models, leading to a Bose glass phase.
Contrary to the Mott insulating and superfluid phases, we find that the Bose glass phase is multifractal.
The same multifractality appears on a Bose glass phase in the periodic square lattice.
Therefore, multifractality is common in a Bose glass phase irrespective of the periodicity of systems.
}

\begin{document}
\maketitle

\section{Introduction}

A quasicrystal retains a structure that is not periodic.  Thus spatial distribution of physical quantities in quasicrystals is nonuniform unlike periodic crystals.
A quasicrystal also has a structure that is not random but ordered, since the vertices of tiles forming a quasicrystalline pattern have a long-range order with sharp Bragg peaks.
As a result, the spatial pattern of physical quantities in quasicrystals is neither random nor uniform~\cite{Araujo_2019,Koga_2020,Fukushima_2023_Nov,Hori_2024_Jan}.
Multifractal and hyperuniform analyses have been proposed in quantitatively characterizing the spatial distribution of vertices~\cite{Halsey1986,Torquato2003,Koga_2024,Bialus_2024_June} and nonuniform spatial patterns of physical quantities including electronic states in quasicrystals and  crystals~\cite{Tokihiro_1998_Sep, Subramaniam_2006, Obuse_2007,Torquato_2016,Mace_2017_Jul,Torquato_2018_June, Fuchs_2019_Sep, Sakai2022, Sakai2022_2,Hamanaka_2024_Jan,Sun_2024}.
So far, it is known that the spatial distribution of lattice points in crystals and vertices in most quasicrystals is hyperuniform but that in random systems is multifractal.

In contrast to states in fermion systems, multifractality and hyperuniformity in boson systems have not been elucidated, although bosonic quasicrystalline systems have been examined experimentally~\cite{Sbroscia_2020_Nov,yu_2022_thesis,Gottlob_2023_Apr, yu_2023} and theoretically~\cite{Ghadimi2020,Gautier_2021,Ciardi_2023} by using ultracold atoms in optical quasicrystals.
Such boson systems can be described by a Bose-Hubbard model, which exhibits Mott insulating phase (superfluid phase) when on-site Coulomb interaction (hopping interaction) is dominant~\cite{Ghadimi2020}.
It has not been clarified whether spatial distribution of physical quantities in those phases exhibits either hyperuniform or multifractal.

Very recently, a Bose glass phase~\cite{Fisher_1989} has been observed in an optical quasicrystal~\cite{yu_2022_thesis,yu_2023}.
In the Bose glass phase, spatial correlation in physical quantities is weak and almost random.
If randomness would determine spatial properties, this Bose glass phase might show a multifractal feature similar to a fermionic random system~\cite{Puschmann_2015, Jack_2021}.
If so, a transition between states to be hyperuniform and multifractal may be possible by controlling disorder in a bosonic quasicrystalline system.

In this paper, we apply the multifractal and hyperuniform analyses to physical quantities in Bose-Hubbard models with and without disorder to clarify similarities and differences between crystalline and quasicrystalline systems.
We find that the system without disorder is hyperuniform in both crystalline and quasicrystalline systems.
The order metric, which quantifies the complexity of the system, becomes more significantly larger as approaching a phase boundary in quasicrystals than in crystals.
The origin of this enhancement in quasicrystals is attributed to a wider variety of vertices which allows a wider variety of the distribution of a physical quantity than in crystals.
Near a transition point, we find a unique feature of superfluid phases where spatial patterns of physical quantities are governed by quasiperiodicity.
We show that one can quantify how complex the distribution of a physical quantity is in comparison with the point distribution by using hyperuniformity, and distinguish crystalline and quasicrystalline systems.
With disorder, we find a hyperuniform Mott insulating phase, a multifractal Bose glass phase, and a hyperuniform superfluid phase.
The presence of disorder makes it difficult to clarify differences between crystalline and quasicrystalline systems.

This paper is organized as follows.
We introduce a mean-field Bose-Hubbard model, perpendicular space, multifractality, hyperuniformity, replica overlap for Bose glass, and the inverse participation ratio (IPR) in Sec.~\ref{sec:model}.
In Sec.~\ref{sec:ep0}, we perform multifractal and hyperuniform analyses for physical quantities without disorder.
The effect of disorder is discussed in Sec.~\ref{sec:withDis}.
Finally, we give a summary of this work in Sec.~\ref{sec:conclusion}.

\section{Model and Method}
\label{sec:model}
\subsection{Mean-field Bose-Hubbard model}
We consider a Bose-Hubbard model with an on-site disorder potential.
In quasicrystals, we use a vertex model, where the links connecting vertices are of the same length and hopping of bosons is along the links.
A mean-field Bose-Hubbard model reads~\cite{Ghadimi2020}
\begin{align}
    H_{\rm MF}=\sum_i H_i + E_0,
    \label{eq:modelBH}
\end{align}
where a local Hamiltonian at an $i$th vertex is given by
\begin{align}
    H_{i}=-J(\psi^{*}_i\hat{b}_i+{\rm H.c.})-(\mu - \epsilon_i) \hat{n}_i+\frac{U}{2}\hat{n}_i(\hat{n}_i -1),
    \label{eq:selfConsistent}
\end{align}
and a constant energy $E_0$ is given by
\begin{align}
E_0=J\sum_{i}\psi_i^*\braket{\hat{b}_i}.
    \label{eq:selfConsistentE0}
\end{align}
Here, $\hat{b}^\dagger_i$ ($\hat{b}_i$) is the creation (annihilation) operator of a boson at the $i$th vertex, $\braket{\hat{b}_i}$ is the local superfluid amplitude, $\psi_i=\sum_{j_i} \braket{\hat{b}_{j_i}}$ where $j_i$ represents a vertex connected to the $i$th vertex by a link with hopping $J$, $\mu$ is the chemical potential, $\epsilon_i$ represents the vertex-dependent random on-site potential, $\hat{n}_i=\hat{b}^\dagger_i\hat{b}_i$, and $U$ denotes an on-site Coulomb interaction.
In the following, we set $U$ as an energy unit, and we take $\mu/U=0.5$ and 1.0.

We perform self-consistent calculations~\cite{Pai2012,Kurdestany2012,Ghadimi2020} for Eq.~(\ref{eq:selfConsistent}) to obtain $\braket{\hat{b}_i}$ and $\braket{\hat{n}_i}$.
Using an initial guess $\psi_i$ and diagonalizing the local Hamiltonian (2), we obtain a wave function of the ground state and calculate $\braket{\hat{n}_i}$ and $\braket{\hat{b}_i}$ in the local Hilbert space that satisfies $\braket{\hat{n}_i}\le n_b$ bosons.
We repeatedly calculate $\psi_i$ by using the obtained $\braket{\hat{n}_i}$ and $\braket{\hat{b}_i}$ until the calculation reaches its convergence.
In the following, we set $n_b=10$.

Some of the present authors have shown that our model in Eq.~(\ref{eq:modelBH}) without disorder, namely $\epsilon_i =0$, realizes the Mott insulating and superfluid phases not only in crystals but also in quasicrystals~\cite{Ghadimi2020}.
We note that $\braket{\hat{b}_i}$ is zero at all vertices (nonzero at some vertices) in the Mott insulating (superfluid) phase.

\subsection{Perpendicular space}

\begin{figure}
	\includegraphics[width=8.6cm]{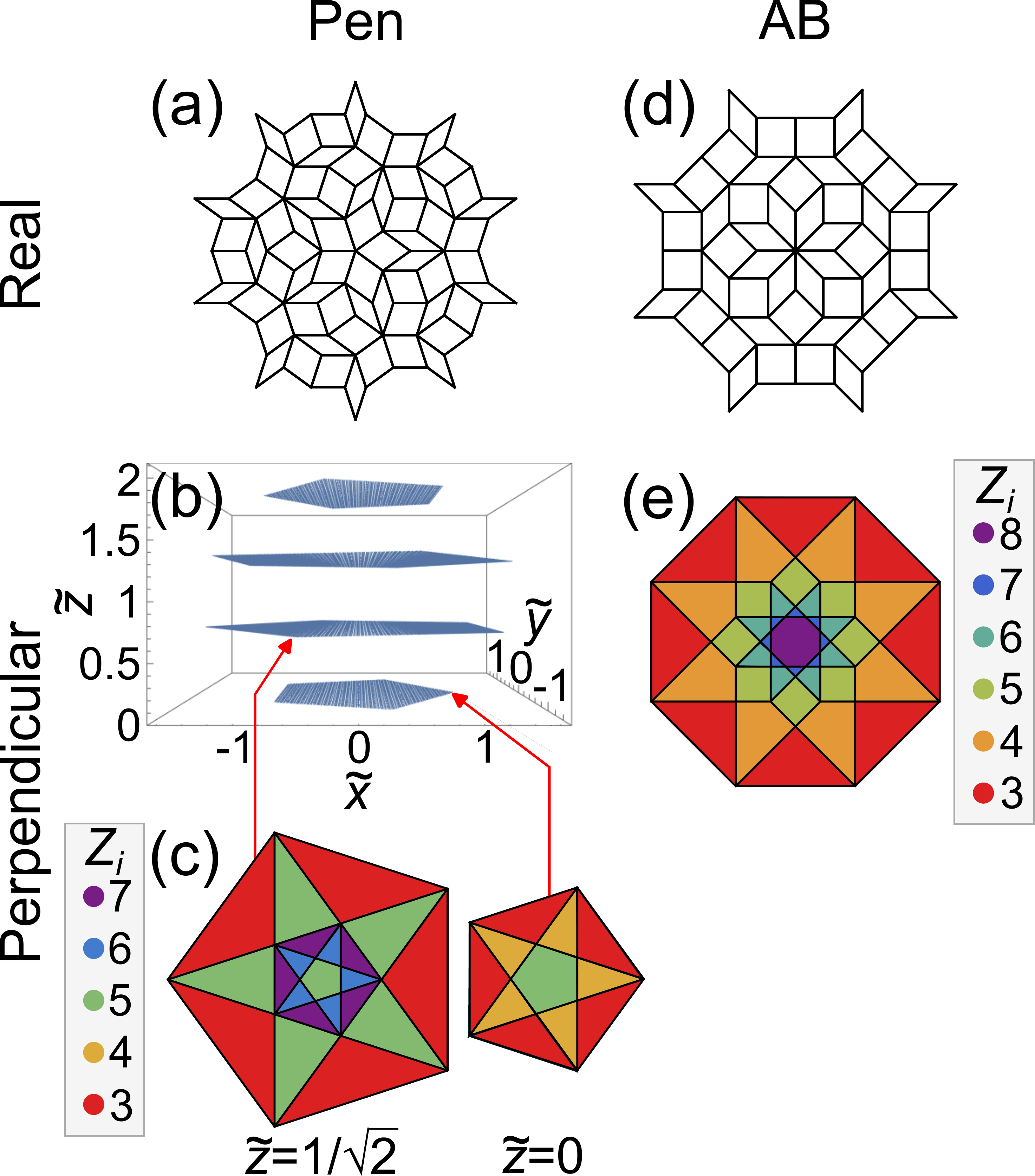}
	\caption{\label{fig:PenAB_real_perp}
 (Color online)
 (a) The real-space representation of Penrose quasicrystals (3rd generation, $N=86$).
 (b) The perpendicular-space representation of Penrose quasicrystals.  
 (c) Coordination number for $\tilde{z}=1/\sqrt{2}$ and 0.
 (d) Same as (a) but for Ammann-Beenker quasicrystals (2nd generation, $N=65$).
 (e) Same as (c) but for Ammann-Beenker quasicrystals. 
 }
\end{figure}

We consider the Bose-Hubbard model (\ref{eq:selfConsistent}) on a Penrose tiling that is a covering of the plane by non-overlapping rhombi as shown in Fig.~\ref{fig:PenAB_real_perp}(a).
Since a Penrose tiling forms a quasicrystalline structure, we call structures formed by a Penrose titling Penrose quasicrystals in this paper.
The vertices of these quasicrystals are obtained by projection from a five-dimensional hypercubic lattice onto a two-dimensional plane.
For the direction perpendicular to this projection, we consider another projection from a five-dimensional hypercubic lattice onto a three-dimensional space, which results in the perpendicular-space representation as shown in Fig.~\ref{fig:PenAB_real_perp}(b).
Any vertex is located at one of the four pentagonal planes at $\tilde{z}=0,$ $1/\sqrt{2},$ $2/\sqrt{2},$ and $3/\sqrt{2}$.
We show two of the four planes for $\tilde{z}=1/\sqrt{2}$ (left panel) and 0 (right panel) in Fig.~\ref{fig:PenAB_real_perp}(c), where the colors distinguish a coordination number $Z_i$ defined as the number of vertices that are connected with the $i$th vertex along links.
The black lines separate vertices with different $Z_i$, resulting in the formation of star shapes in the pentagonal planes.
The average of $Z_i$ for the entire system is $\overline{Z_i}=4$ in Penrose quasicrystals, which agrees with that of square lattice.
Note that the pentagonal planes at $\tilde{z}=3/\sqrt{2}$ and $\tilde{z}=2/\sqrt{2}$ are identical to the planes at $\tilde{z}=0$ and $\tilde{z}=1/\sqrt{2}$, respectively, with reversing them upside down.
This allows us to focus on only two planes at $\tilde{z}=0$ and $1/\sqrt{2}$.

We also consider the Bose-Hubbard model (\ref{eq:selfConsistent}) on an Ammann-Beenker tiling as shown in Fig.~\ref{fig:PenAB_real_perp}(d).
We call structures with an Ammann-Beenker tiling Ammann-Beenker quasicrystals, whose corresponding hypercubic lattice is in a four-dimensional space.
As shown in Fig.~\ref{fig:PenAB_real_perp}(e), its perpendicular-space representation forms a two-dimensional octagon.
Note that the coordination number $Z_i$ increases with approaching the center of the octagonal plane.
The averaged coordination number is the same as Penrose quasicrystals, namely $\overline{Z_i}=4$.

We can examine how quasiperiodicity affects the spatial distribution of a physical quantity by plotting it in the perpendicular-space representation.
If the distribution is similar to Fig.~\ref{fig:PenAB_real_perp}(c) [\ref{fig:PenAB_real_perp}(e)] in Penrose [Ammann-Beenker] quasicrystals, one can conclude that the physical quantity is governed by local environment of vertices that produces quasiperiodicity.
If not, one can conclude that the physical quantity is not significantly affected by quasiperiodicity.

In the following, we use the total number of vertices $N=75806$ (10th generation) and $N=65281$ (6th generation) for Penrose and Ammann-Beenker quasicrystals, respectively (see Appendix).
For comparison, we also use square lattice with $N=90000$.
We use open boundary conditions.


\subsection{Multifractality}

For clarifying the effect of quasiperiodicity on physical quantities such as $\braket{\hat{b}_i}$ or $\braket{\hat{n}_i}$, we employ the multifractal analysis~\cite{Halsey1986}.
The multifractal analysis is one of the methods to quantify the fractal property of a nonuniform spatial pattern, which has been employed also in quasicrystals~\cite{Sakai2022, Sakai2022_2}.

The definition of multifractality depends on a spatial dimension.
In this paper, we concentrate on the multifractality in two dimensions.
We introduce a circular window of radius $R$ and consider an $R$-dependent quantity given by
\begin{align}
    D(\{n_i\},q,R)=\frac{1}{1-q} \frac{\ln{\frac{\sum_i |n_i|^q}{(\sum_i |n_i|)^q}}}{\ln \sqrt{N_{R}}},
    \label{eq:RDepMFD}
\end{align}
where $\{n_i\}$ is the distribution of physical quantity $n_i$ and the summation of $i$ is taken for all vertices within a circular window of radius $R$, $N_R$ is the number of vertices within the circular window, and $q$ is a real variable.
By taking a limit $R\rightarrow \infty$ for Eq.~(\ref{eq:RDepMFD}), the $q$-dependent multifractal dimension is defined as
\begin{align}
    D^{\infty}(\{n_i\},q)=\lim_{R\rightarrow \infty}D(\{n_i\},q,R).
    \label{eq:RLimMFD}
\end{align}
This dimension for a negatively (positively) large $q$ represents a fractal property of larger (smaller) values of $n_i$.

A given system is multifractal in terms of $\{n_i\}$ if and only if $D^{\infty}(\{n_i\},q)\neq 2$ for a certain $q$.
In general, $D^{\infty}(\{n_i\},q)$ is a decreasing function of $q$, and thus it is enough to calculate $D^{\infty}(\{n_i\},q)$ with a large $q$ and a small $q$ to judge if a system is multifractal or not.
In the following, we concentrate on $q=\pm 10$.
We define multifractalness as
\begin{equation}
\begin{split}
    &{\rm multifractalness~of~}\{n_i\}\\
    &= D^{\infty}(\{n_i\},q=-10)-D^{\infty}(\{n_i\},q=10).    
\end{split}
\end{equation}
This quantifies a tendency to be multifractal for a given system.
That is, when multifractalness is larger than 0, the system is multifractal.

\subsection{Hyperuniformity}

Hyperuniform analysis~\cite{Torquato2003,Torquato_2016,Torquato_2018_June} is also one of the methods to quantify a property of a nonuniform spatial pattern.
Hyperuniform analysis has been used for the distribution of physical quantities in quasicrystals~\cite{Fuchs_2019_Sep, Sakai2022, Sakai2022_2}.
In the following, we explain a hyperuniformity in two-dimensional systems.
First, the sum of a physical quantity $n_i$ inside a circular window of radius $R$ is given by
\begin{align}
    N(R) = \sum_i n_i \Theta(R-|\bm{r}_i-\bm{r}_c|),
    \label{eq:sumCircle}
\end{align}
where $\Theta(\bullet)$ is the Heaviside step function, $\bm{r}_i$ is a position of the $i$th vertex, and $\bm{r}_c$ represents the center of the circular window.
The variance of $N(R)$ with respect to the random sampling of $\bm{r}_c$ is given by 
\begin{align}
    \sigma(R)^2 = \left[N(R)^2\right]_{\bm{r}_c} - \left(\left[N(R)\right]_{\bm{r}_c}\right)^2,
    \label{eq:variance}
\end{align}
where $\left[\bullet\right]_{\bm{r}_c}$ denotes the average of $\bullet$ with respect to the random sampling of $\bm{r}_c$.

If the variance $\sigma(R)^2$ scales slower than the area of the window, a given system is hyperuniform in terms of $\{n_i\}$: the condition to be hyperuniform is that, defining 
\begin{align}
    A(R)=\frac{\sigma(R)^2}{(R/D)^2},
    \label{eq:AR}
\end{align}
$\lim_{R\rightarrow\infty} A(R)=0$, where $D$ is the length of links.
In general, there are three classes of hyperuniformity~\cite{Torquato2003} that are classified according to the $R$ dependence of $\sigma(R)^{2}$ in two dimensions as 
\begin{align}
    \sigma(R)^{2}\sim\left\{
\begin{array}{lll}
R & {\rm Class~I}\\
R \ln R & {\rm Class~I\hspace{-1.2pt}I}\\
R^{2-\alpha} (0<\alpha<1) & {\rm Class~I\hspace{-1.2pt}I\hspace{-1.2pt}I}.\\
\end{array}
\right.
    \label{eq:classHyper}
\end{align}
For the point distribution where $n_{i}=1$, the system is Class-I hyperuniform in all crystals and most quasicrystals~\cite{Torquato2003}.

When a system is Class-I hyperuniform, a proportionality coefficient in front of $R$ in Eq.~(\ref{eq:classHyper}), called order metric, controls the complexity of the distribution of $n_i$.
Defining 
\begin{equation}
    B(R) = \frac{\sigma(R)^2}{R/D},
    \label{eq:BR}
\end{equation}
we numerically obtain the order metric as an average of $B(R)$ at large distance.
The order metrics of the point distribution in most quasicrystals are larger than those in most crystals, due to the complex structures of quasicrystals.

In most cases, a quasicrystalline system is either multifractal or hyperuniform for each physical quantity.
Note that multifractal systems are never hyperuniform, and vice versa.
Hence, as we confirm that a system is multifractal (hyperuniform), we immediately conclude that the system is not hyperuniform (multifractal).

\subsection{Replica overlap for Bose glass}

We consider an $L\times L$ matrix $\mathbf{Q}=(q^{\alpha\beta})_{1\le \alpha, \beta \le L}$ whose $(\alpha,\beta)$ component is given by
\begin{align}
    q^{\alpha\beta}=\frac{1}{N}\sum_{i=1}^N \kappa_i^{\alpha} \kappa_i^{\beta},
    \label{eq:qAlphaBeta}
\end{align}
where $N$ is the total number of vertices, $\alpha$ and $\beta$ represent replica indices for a shot of randomness, $L$ is the number of kinds of $\alpha$ and $\beta$, and a compressibility $\kappa^\alpha_i=\braket{\hat{n}_i^2}-\braket{\hat{n}_i}^2$ denotes the fluctuation of vertex-dependent boson density $\braket{\hat{n}_i}$ for a replica index $\alpha$.
In the absence of disorder, the superscript $\alpha$ denoting the replica number is omitted, i.e., $\kappa_i$, in the following.
The quantity $q^{\alpha\beta}$ represents the overlap of the fluctuations of $\braket{\hat{n}_i}$ between the two shots of randomness $\alpha$ and $\beta$.
Note that $q^{\alpha\beta}$ is similar to the Edwards-Anderson order parameter for spin glass~\cite{Edwards_1975,Parisi_1983} and Bose glass~\cite{Khellil_2016,Thomson_2016}.

We calculate the fluctuation of $q^{\alpha\beta}$ as an indicator for the Bose glass phase.
All of information on the overlaps of $\kappa_i^{\alpha}$ is contained in a set of upper-triangular elements of $\mathbf{Q}$ excluding diagonal elements.
By using this set of $q^{\alpha \beta}$'s, we define a replica overlap $q_{\rm BG}$ as 
\begin{align}
q_{\rm BG}=\sqrt{
\frac{\frac{\sum_{\alpha<\beta}(q^{\alpha\beta})^2}{\frac{L(L-1)}{2}}
    -\left(\frac{\sum_{\alpha<\beta}q^{\alpha\beta}}{\frac{L(L-1)}{2}}\right)^2}
    {
    \left(\frac{\sum_{\alpha<\beta}q^{\alpha\beta}}{\frac{L(L-1)}{2}}\right)^2
    }
    },
    \label{eq:qBG13}
\end{align}
where we introduce a factor $L(L-1)/2$ to remove a trivial effect from the number of types of randomness.

\subsection{Inverse participation ratio (IPR)}

IPR quantifies the degree of localization of a nonuniform spatial pattern, which has been applied in quasicrystals~\cite{Araujo_2019,Ghadimi2023}. 
We define IPR for a set of physical quantities $\{n_i\}_{i=1,2,\ldots,N}$ as
\begin{align}
    {\rm IPR \hspace{0.1cm} of \hspace{0.1cm}} \{n_i\} = \sum_{i=1}^{N} \left( \frac{|n_i|}{\sum_{j=1}^{N}|n_j|} \right)^2.
    \label{eq:IPR}
\end{align}
Since the quantity in the summation of $i$, $|n_i|/\sum_j|n_j|$ is normalized, IPR does not depend on the magnitudes of $n_i$.
We always obtain $1/N \le {\rm IPR} \le 1$, and IPR is smaller (larger) when the physical quantity is delocalized  (localized) in a system.

\section{Without Disorder}
\label{sec:ep0}
In this section, we consider a model without disorder, $\epsilon_i =0$, for $\mu/U=0.5$ and 1.0.
Figure~\ref{fig:orderMetric}(a) shows a schematic phase diagram of our model in Eq.~(\ref{eq:modelBH})~\cite{Ghadimi2020}.
The horizontal axis represents a hopping amplitude $\overline{Z_i} J/U$, and the upper (lower) panel shows a phase diagram for $\mu/U=0.5$ (1.0), where MI and SF denote the Mott insulating phase and superfluid phase, respectively.
The system is in the Mott insulating phase when $\braket{\hat{b}_i}=0$, $\braket{\hat{n}_i}$ is equal integer number, and $\kappa_i=0$ for all vertices.
On the other hand, the system is in the superfluid phase when $\braket{\hat{b}_i}\neq0$, $\braket{\hat{n}_i}\neq1$, and $\kappa_i\neq 0$ for a certain vertex.
When $\mu/U=0.5$, the system undergoes a phase transition from the Mott insulating phase with $\braket{\hat{n}_i}=1$ to the superfluid phase at the critical hopping strength $J=J_{\rm c}$.
The values of $\overline{Z_i}J_{\rm c}/U$ in both Penrose and Ammann-Beenker quasicrystals are approximately 0.16, which is smaller than that of square lattice being 0.17~\cite{Ghadimi2020}.
When $\mu/U=1.0$, the system is in the superfluid phase regardless of the strength of hopping $J$.

\subsection{Hyperuniformity}
\label{subsec:Hyp}

\begin{figure}
	\includegraphics[width=8.6cm]{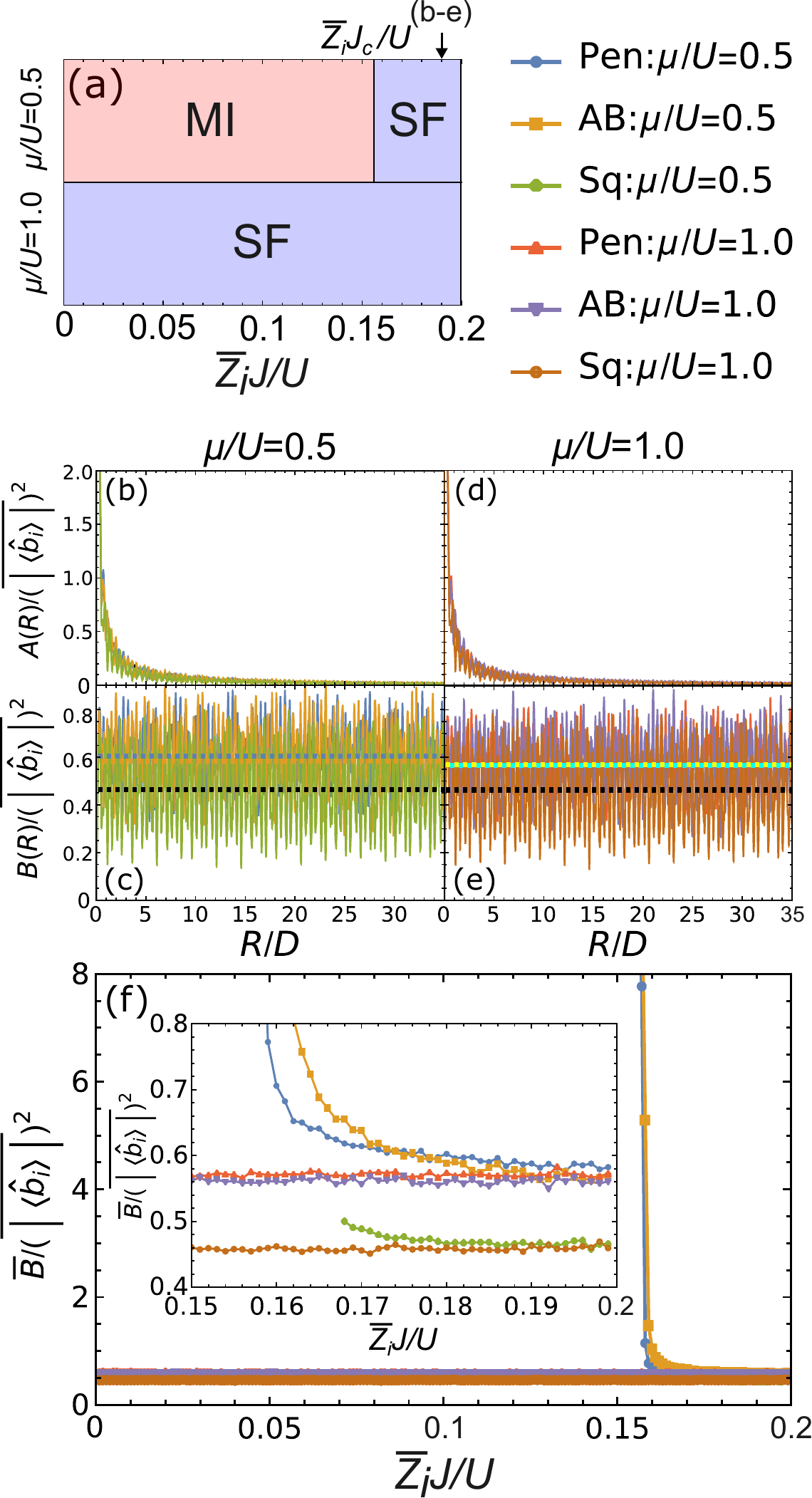}
	\caption{\label{fig:orderMetric}
 (Color online)
(a) Schematic phase diagram of our model in Eq.~(\ref{eq:modelBH}).
(b) $A(R)$ as a function of $R/D$ for $\overline{Z_i}J/U=0.19$ and $\mu/U=0.5$.
(c) Same as (b) but for $B(R)$.
(d) Same as (b) but for $\mu/U=1.0$.
(e) Same as (c) but for $\mu/U=1.0$.
Horizontal colored lines in (c) and (e) represent an average value of $B(R)$ for each system.
(f) Order metric $\overline{B}$ as a function of hopping amplitude for both $\mu/U=0.5$ and 1.0. The inset shows behaviors near the phase boundary shown in (a). Sq represents square lattice and Pen (AB) represents Penrose (Ammann-Beenker) quasicrystals, and MI (SF) denotes the Mott insulating (superfluid) phase.
}
\end{figure}

By solving Eq.~(\ref{eq:modelBH}) self-consistently, we obtain the distribution of $\braket{\hat{b}_i}$, $\braket{\hat{n}_i}$, and $\kappa_i$.
To clarify how quasiperiodicity of a system affects the distribution of physical quantities, we apply a hyperuniform analysis to self-consistently obtained physical quantities.
We focus on hyperuniformity in $\left|\braket{\hat{b}_i}\right|$.
In the Mott insulating phase, the system is hyperuniform because $\braket{\hat{b}_i}=0$.
For the superfluid phase, we show $A(R)$ in Eq.~(\ref{eq:AR}) normalized by $\left(\overline{|\braket{\hat{b}_i}|}\right)^2$ as a function of $R/D$ for $\mu/U=0.5$ and $\overline{Z_i} J/U=0.19$ in Fig.~\ref{fig:orderMetric}(b).
As indicated by a downward black arrow at the upper-right corner in Fig.~\ref{fig:orderMetric}(a), the system is superfluid for this parameter set.
The blue, orange, and green lines represent the results for Penrose quasicrystal, Ammann-Beenker quasicrystal, and square lattice, respectively.
All systems are hyperuniform because $A(R)$ approaches zero in the limit of $R\rightarrow \infty$.
We also find that the systems are hyperuniform regardless of hopping amplitude (not shown).
Figure~\ref{fig:orderMetric}(c) is the same as Fig.~\ref{fig:orderMetric}(b) but for $B(R)$ in Eq.~(\ref{eq:BR}).
The blue, orange, and horizontal black-dashed lines in Fig.~\ref{fig:orderMetric}(c) represent $\overline{B}/\left(\overline{\left|\braket{\hat{b}_i}\right|}\right)^2$ for Penrose quasicrystal, Ammann-Beenker quasicrystal, and square lattice, respectively, where $\overline{B}$ is the average of $B(R)$ within $30\le R/D \le 35$, which we take as order metric.
As shown in Fig.~\ref{fig:orderMetric}(c), $B(R)$ is fluctuating around $\overline{B}$.
This indicates Class~I in hyperuniformity.

Figure~\ref{fig:orderMetric}(d) is the same as Fig.~\ref{fig:orderMetric}(b) but for $\mu/U=1.0$.
The red, purple, and brown lines represent the results for Penrose quasicrystal, Ammann-Beenker quasicrystal, and square lattice, respectively.
Figure~\ref{fig:orderMetric}(e) is the same as Fig.~\ref{fig:orderMetric}(c) but for $\mu/U=1.0$.
The yellow, cyan, and horizontal black-dashed lines in Fig.~\ref{fig:orderMetric}(e) represent $\overline{B}/\left(\overline{|\braket{\hat{b}_i}|}\right)^2$ for Penrose quasicrystal, Ammann-Beenker quasicrystal, and square lattice, respectively.
Similar to the case of $\mu/U=0.5$, Fig.~\ref{fig:orderMetric}(e) indicates Class~I in hyperuniformity for all systems.

Figure~\ref{fig:orderMetric}(f) shows the order metric $\overline{B}$ as a function of hopping strength $J$. 
The inset is a magnified one near the phase boundary.
According to our calculations, the order metric of the point distribution is 0.58 in Penrose quasicrystals, 0.57 in Ammann-Beenker quasicrystals, and 0.46 in square lattice~\cite{Torquato2003,Koga_2024}.
In Fig.~\ref{fig:orderMetric}(f), for $\mu/U=1.0$ or large $J$ for $\mu/U=0.5$, the normalized $\overline{B}$ is close to the order metric of the point distribution.
This can be understood as follows.
Since we divide $\overline{B}$ by $\left(\overline{|\braket{\hat{b}_i}|}\right)^2$ in Fig.~\ref{fig:orderMetric}, we can regard $n_i$ in Eq.~(\ref{eq:sumCircle}) as $\left|\braket{\hat{b}_i}\right|/\overline{|\braket{\hat{b}_i}|}$.
This leads to $\left|\braket{\hat{b}_i}\right|/\overline{|\braket{\hat{b}_i}|}=1$ in square lattice if there is no fluctuation of $\left|\braket{\hat{b}_i}\right|$, resulting in the order metric same as that of the point distribution.
In quasicrystals, $\left|\braket{\hat{b}_i}\right|$ at large $J$ only depends on $Z_i$.
As $J$ increases, $\overline{|\braket{\hat{b}_i}|}$ increases and the variation of $|\braket{\hat{b}_i}|$ becomes relatively small.
This explains that the order metric of $\left|\braket{\hat{b}_i}\right|/\overline{|\braket{\hat{b}_i}|}$ is close to that of the point distribution when $J$ is large.
Note that, for $\mu/U=1.0$, we obtain a uniform distribution regardless of a value of $J$.

As $J$ approaches a transition point, a typical correlation length increases~\cite{Ghadimi2020}, which results in a wider variety of the distribution of a physical quantity.
In crystals (quasicrystals), a width of this variety has (does not have) a maximum limit since the number of distinct vertices is finite (infinite).
It follows that we observe a more significant increase in the order metric in quasicrystals than in crystals.
This indicates that the order metric defined by the concept of hyperuniformity is a useful concept to differentiate between crystalline and quasicrystalline bosonic systems.
Furthermore, we consider that one should pay attention to a difference between the order metric of a physical quantity and that of the point distribution.
It clarifies if the distribution of a physical quantity reflects the point distribution or not.
In addition, it quantifies how complex the distribution is in comparison with the point distribution, as exampled in Fig.~\ref{fig:orderMetric}(f).

\subsection{Near phase boundary}
\label{subsec:PB}

We focus on a phase boundary between the Mott insulating and superfluid phases.
At the vicinity of the phase boundary, the system may show an exotic phase such as Bose glass~\cite{Niederle2015,Johnstone2021,Ciardi_2023}.
The system is in the Bose glass phase when superfluid vertices are randomly distributed.
On the other hand, the system is in the superfluid phase when the fraction of superfluid vertices is large.
We note that the Bose glass phase has been observed in a system of ultracold atoms in optical 
quasicrystals~\cite{yu_2022_thesis,yu_2023}.

To clarify whether an exotic phase appears, we introduce a measure of superfluidity at each vertex~\cite{Niederle2013} as
\begin{align}
    S_i=\left\{
\begin{array}{ll}
0 & {\rm if}\quad 1-\gamma_n \le \braket{\hat{n}_i} \le 1+\gamma_n, \\
1 & {\rm else},
\end{array}
\right.
    \label{eq:SiDef}
\end{align}
where $\gamma_n$ is a small number.
In the following, we set $\gamma_n=5\times10^{-3}$.
We call the $i$th vertex with $S_i=1$ (0) a superfluid (Mott insulating) vertex.
Note that $S_i=1$ (0) for all vertices in the superfluid (Mott insulating) phase.
Since the Mott insulating phase changes to the superfluid phase with increasing $J$, there would be some signatures in the distribution of $S_i$ at a certain value of $J$ if an exotic phase appears.
To check this, we search a critical hopping amplitude $J_i^{\rm c}$ for each vertex where $S_i$ changes from 0 (Mott insulating vertex) to 1 (superfluid vertex).

\subsubsection{Penrose quasicrystals}

\begin{figure}
	\includegraphics[width=8.4cm]{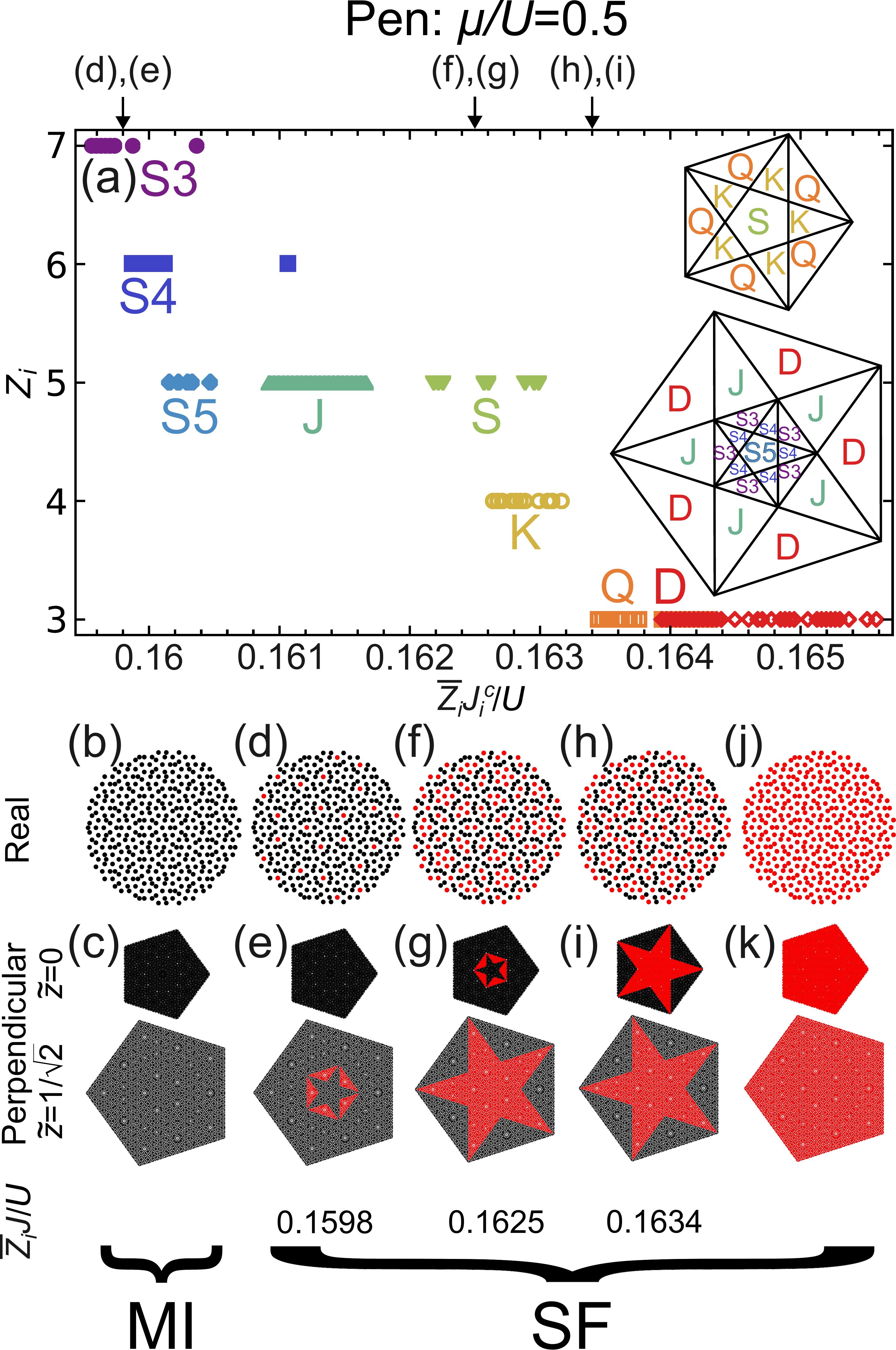}
	\caption{\label{fig:MISFboundarySi_Pen}
(Color online)
(a) The distribution of $J_i^{\rm c}$ for a given $Z_i$ as a function $\overline{Z_i} J/U$ for a Penrose quasicrystal with $\mu/U=0.5$.
Symbols are assigned for each group of $J_i^{\rm c}$ and put into corresponding regions in the $\tilde{z}=0$ and $1/\sqrt{2}$ perpendicular-space planes in the inset.
The distribution of $S_i$ in the real space and perpendicular space for [(b) and (c)] the Mott insulating phase, [(d) and (e)] $\overline{Z_i} J/U=0.1598$, [(f) and (g)] $\overline{Z_i} J/U=0.1625$, [(h) and (i)] $\overline{Z_i} J/U=0.1634$, [(j) and (k)] the superfluid phase, respectively.
}
\end{figure}

First, we discuss Penrose quasicrystals.
After a self-consistent calculation with the open boundary condition for $N=75806$, we remove vertices at the edges in order to eliminate the effect of boundaries (see Appendix).
We take 55651 vertices within $R/D\le 120$ and calculate $J_i^{\rm c}$ for each vertex.
Then the values of $J_i^{\rm c}$ are grouped together by the coordination number $Z_i$ and are plotted in Fig.~\ref{fig:MISFboundarySi_Pen}(a) as colored symbols.
We find that, the larger $Z_i$ is, the smaller $J_i^{\rm c}$ is.
For each $Z_i$, they group around certain values of $J$. 
Each group is labeled by symbols such as S3 and S4~\cite{deBrujin_1981_1,deBrujin_1981_2}.
Such a grouping of $J_i^{\rm c}$ comes from the difference of local environment beyond nearest-neighbor links.
Considering the perpendicular space of Penrose quasicrystals, we can map the symbols onto the $\tilde{z}=0$ and $1/\sqrt{2}$ planes as shown in the inset of Fig.~\ref{fig:MISFboundarySi_Pen}(a).

In the Mott insulating phase where $\overline{Z_i} J/U\lesssim0.159$, all vertices are the Mott insulating vertex ($S_i=0$). 
We show the real-space representation of $S_i=0$ in Fig.~\ref{fig:MISFboundarySi_Pen}(b) by putting black dots on each vertex around a center of a Penrose quasicrystal within $R/D\le 10$.
In the perpendicular space,  all vertices in the $\tilde{z}=0$ and $1/\sqrt{2}$ planes are colored in black, i.e., $S_i=0$ as shown in the upper and lower panels of Fig.~\ref{fig:MISFboundarySi_Pen}(c), respectively.
At $\overline{Z_i} J/U=0.1598$, some of the $Z_i=7$ vertices satisfying $\overline{Z_i}J_i^{\rm c}/U<0.1598$ are the superfluid vertex ($S_i=1$) as expected from Fig.~\ref{fig:MISFboundarySi_Pen}(a).
In this case, the superfluid vertices (red dots) are regularly distributed in the real space as shown in Fig.~\ref{fig:MISFboundarySi_Pen}(d).
In the perpendicular space, a part of the $1/\sqrt{2}$ plane is colored in red as seen in Fig.~\ref{fig:MISFboundarySi_Pen}(e).
With increasing $\overline{Z_i}J_i^{\rm c}/U$ from 0.1625 to 0.1634, the number of red dots increases keeping a regular pattern in Figs.~\ref{fig:MISFboundarySi_Pen}(f) and \ref{fig:MISFboundarySi_Pen}(h).
Accordingly, a star shape colored in red appears in the $\tilde{z}=0$ and $1/\sqrt{2}$ planes as shown in Figs.~\ref{fig:MISFboundarySi_Pen}(g) and \ref{fig:MISFboundarySi_Pen}(i).
In the superfluid phase, all vertices become the superfluid vertex  (red dots) as seen in Figs.~\ref{fig:MISFboundarySi_Pen}(j) and \ref{fig:MISFboundarySi_Pen}(k).
Based on these evolutions of $S_i$ with $J$ near the phase boundary, we conclude that the distribution of the superfluid vertices is not random but regular and thus it reflects the vertex structure of Penrose quasicrystals.
This implies that there is no Bose glass phase near the phase boundary.
We will discuss the Bose glass phase in the presence disorder in Sec.~\ref{sec:withDis}.

\begin{figure}
	\includegraphics[width=8.6cm]{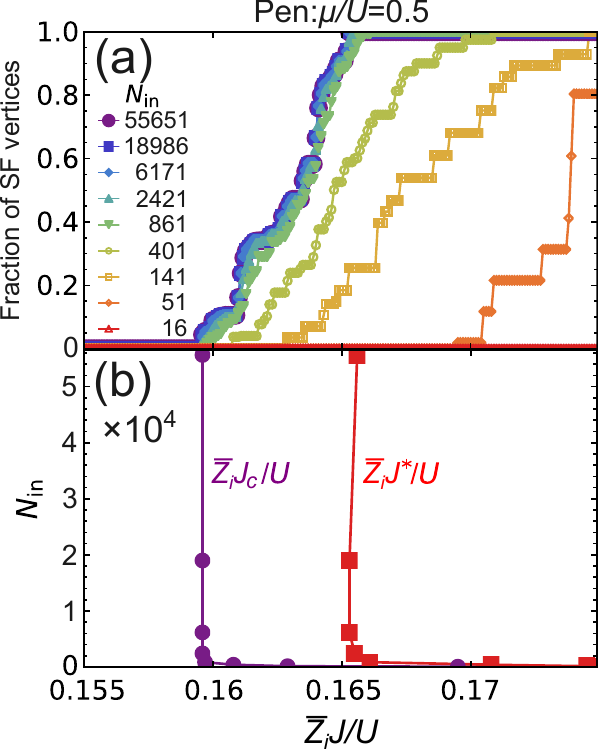}
	\caption{\label{fig:criticalJ_N_combined_Pen}
 (Color online)
(a) The hopping amplitude $J$ dependence of the fraction of superfluid vertices for various size $N_\mathrm{in}$ of Penrose quasicrystals with $\mu/U=0.5$. 
(b) The critical hopping amplitude of $J_{\rm c}$ (purple circles) and $J^{*}$ (red squares) for various $N_\mathrm{in}$.
}
\end{figure}

As shown in Fig.~\ref{fig:orderMetric}(a), we find a significant increase of order metric near the phase boundary only in quasicrystals.
Since order metric measures the complexity of a given system, its increase may be attributed to the presence of nonuniform superfluid phase where some of the vertices are not superfluid vertices seen as discussed above.
To examine whether the nonuniform superfluid phase is caused by a finite-size effect or not, we show in Fig.~\ref{fig:criticalJ_N_combined_Pen}(a) the fraction of superfluid vertices as a function of $J$  near the phase boundary for various system sizes $N_{\rm in}$, where $N_{\rm in}$ is the number of vertices that exclude those around edge.
The fraction is defined as the ratio of superfluid vertices to $N_{\rm in}$.
We find that the fraction is 0 (1) for $J<J_{\rm c}$ ($J^{*}<J$), where $J_{\rm c}$ ($J^{*}$) depends on $N_{\rm in}$, and $J^{*}$ refers to a characteristic $J$.
The nonuniform superfluid phase appears for $J_{\rm c}<J<J^{*}$.
The difference $J^{*}-J_{\rm c}$ is expected to remain finite in the limit of $N_{\rm in}\rightarrow \infty$ as shown Fig.~\ref{fig:criticalJ_N_combined_Pen}(b).
Therefore, the presence of nonuniform superfluid phase is intrinsic in Penrose quasicrystals.
Accordingly, we conclude that the increase of order metric near the phase boundary is also intrinsic.

\subsubsection{Ammann-Beenker quasicrystals}

\begin{figure}
	\includegraphics[width=8.6cm]{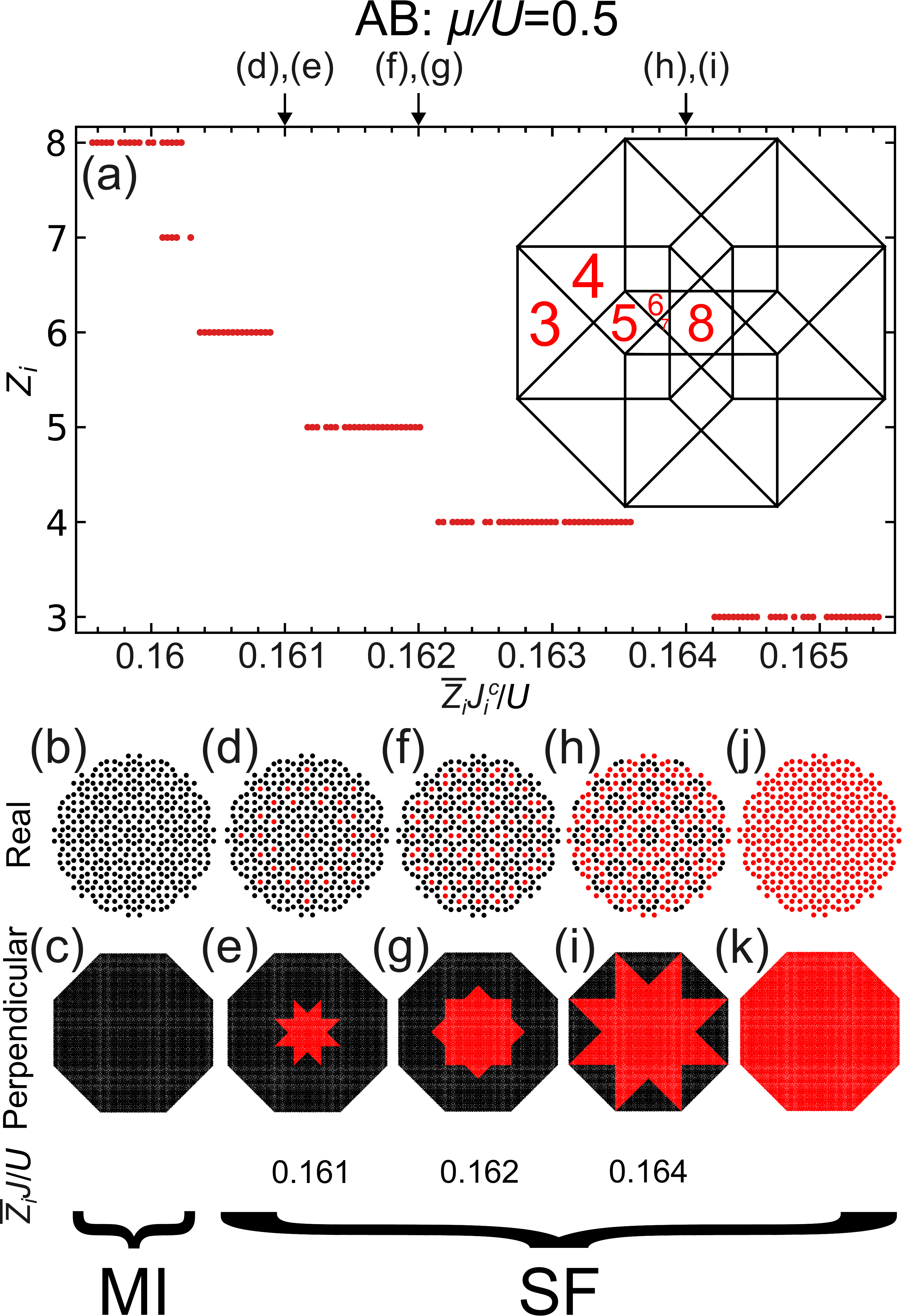}
	\caption{\label{fig:MISFboundarySi}
 (Color online)
 Same as Fig.~\ref{fig:MISFboundarySi_Pen} but for Ammann-Beenker quasicrystals with $\mu/U=0.5$. (a) The distribution of $J_i^{\rm c}$ for a given $Z_i$ as a function $\overline{Z_i} J/U$. The inset presents the entire perpendicular space. The distribution of $S_i$ in the real space [(b) to (j)] and perpendicular space [(c) to (k)]. We use [(d) and (e)] $\overline{Z_i} J/U=0.161$, [(f) and (g)] $0.162$, and [(h) and (i)] $0.164$.
}
\end{figure}

Next, we examine behaviors near the phase boundary in Ammann-Beenker quasicrystals.
After a self-consistent calculation with the open boundary condition for $N=65281$, we use 45873 vertices within $R/D\le 110$ for the evaluation of $S_i$ and $J_i^{\rm c}$.
Figure~\ref{fig:MISFboundarySi}(a) shows $\overline{Z_i} J_i^{\rm c}/U$ for various $Z_i$.
We find that $J_i^{\rm c}$ decreases with increasing $Z_i$, which is the same as Penrose quasicrystals.
The inset represents an octagon in the perpendicular space, where the number of $Z_i$ is assigned only in a one-eighth region of the entire space.

In Fig.~\ref{fig:MISFboundarySi}(b), we show the distribution of $S_i$ around a center of the real space $R/D\le 10$ for the Mott insulating phase ($S_i=0$ represented by black dots).
Figure~\ref{fig:MISFboundarySi}(c) is for the perpendicular space and all vertices are Mott insulating vertices.
Figures~\ref{fig:MISFboundarySi}(d) [\ref{fig:MISFboundarySi}(e)], \ref{fig:MISFboundarySi}(d) [\ref{fig:MISFboundarySi}(e)], and \ref{fig:MISFboundarySi}(d) [\ref{fig:MISFboundarySi}(e)] are the same as Fig.~\ref{fig:MISFboundarySi}(b) [\ref{fig:MISFboundarySi}(c)] but for $\overline{Z_i} J/U=0.161$, 0.162, and 0.164, respectively.
All vertices for $Z_i$ more than 5, 4, and 3 are superfluid vertices ($S_i=1$ represented by red dots) for $\overline{Z_i} J/U=0.161$, 0.162, and 0.164, respectively.
Figures~\ref{fig:MISFboundarySi}(j) and \ref{fig:MISFboundarySi}(k) are for the superfluid phase where all the vertices are $S_i=1$.
Accordingly, a nonuniform superfluid phase reflecting a quasiperiodic structure of the Ammann-Beenker quasicrystals also appears near the phase boundary, which is similar to Penrose quasicrystals.

\begin{figure}
	\includegraphics[width=8.6cm]{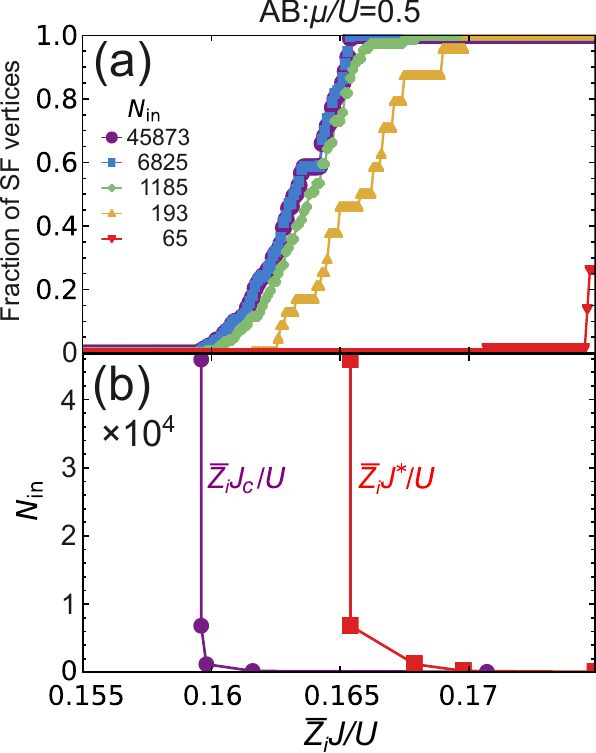}
	\caption{\label{fig:criticalJ_N_combined}
 (Color online)
 Same as Fig.~\ref{fig:criticalJ_N_combined_Pen} but for Ammann-Beenker quasicrystals with $\mu/U=0.5$.
}
\end{figure}

We show in Fig.~\ref{fig:criticalJ_N_combined}(a) the fraction of superfluid vertices as a function of $\overline{Z_i}J/U$ for various $N_{\rm in}$.
We also show $J_{\rm c}$ (purple) and $J^{*}$ (red) in Fig.~\ref{fig:criticalJ_N_combined}(b).
In the limit of $N_{\rm in}\rightarrow \infty$, $J^{*}-J_{\rm c}$ is expected to remain finite.
Accordingly, we conclude that the presence of nonuniform superfluid phase and the resulting increase of order metric are not due to a finite-size effect but intrinsic behaviors.

\subsubsection{Comparison between crystals and quasicrystals}

\begin{figure}[b]
	\includegraphics[width=8.6cm]{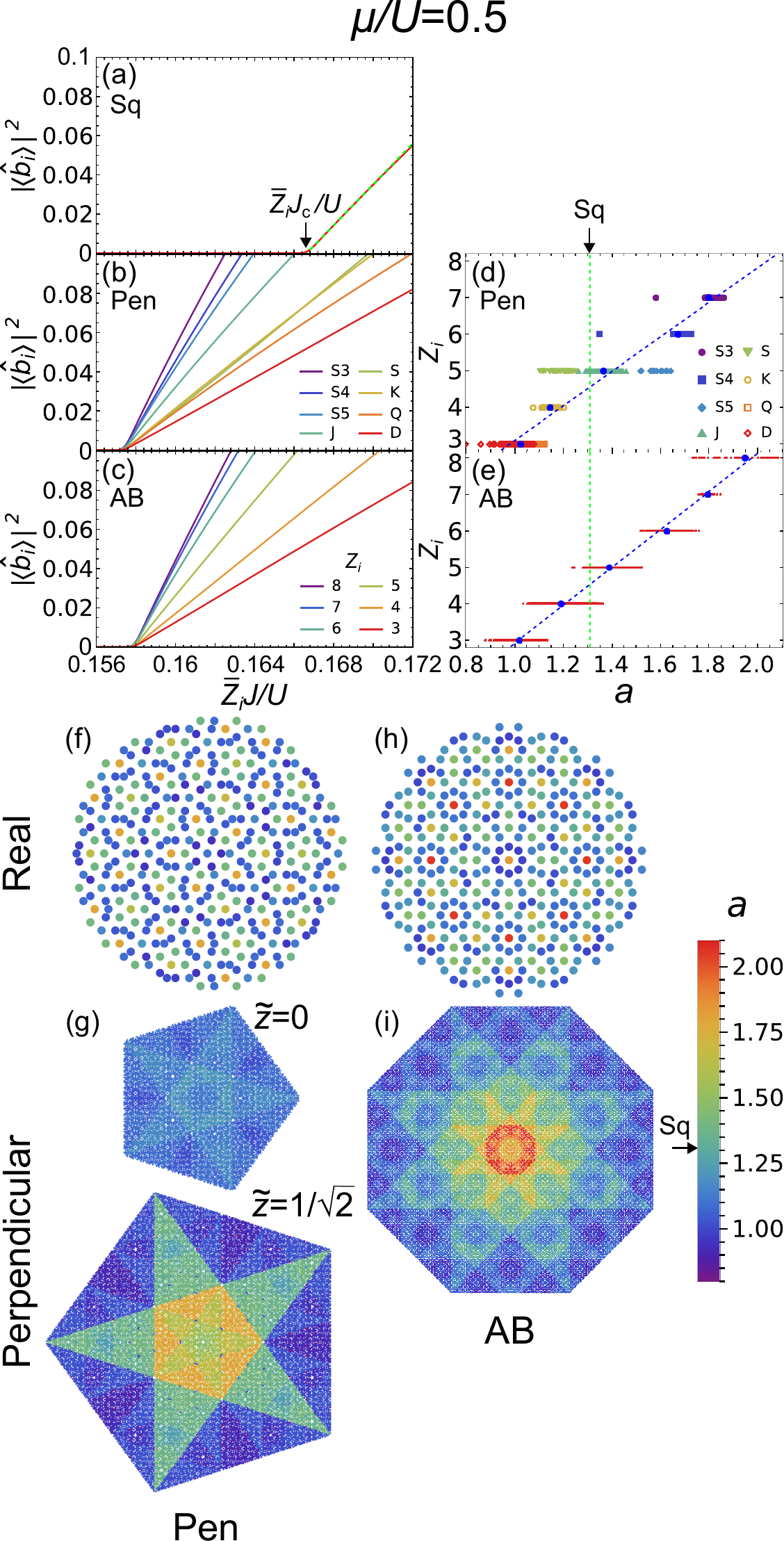}
	\caption{\label{fig:comparisonBeta}
 (Color online)
 (a) $|\braket{\hat{b}_i}|^2$ as a function of $\overline{Z_i}J/U$.
 (b) Same as (a) but for various types of vertices in a Penrose quasicrystal.
 (c) Same as (b) but for an Ammann-Beenker quasicrystal.
 (d) The relationship between $Z_i$ and $a$ in a Penrose quasicrystal.
 (e) Same as (d) but for an Ammann-Beenker quasicrystal.
 The symbols in (b) and (d) denote groups of $Z_i$ defined in Fig.~\ref{fig:MISFboundarySi_Pen}(a).
 The vertical green-dotted line in (d) and (e) represents the value of $a$ in a square lattice.
 (f) The real-space distribution of $a$ in a Penrose quasicrystal.
 (g) The perpendicular-space distribution of $a$ in a Penrose quasicrystal for $\tilde{z}=0$ (upper panel) and $1/\sqrt{2}$ (lower panel).
 (h) Same as (f) but for an Ammann-Beenker quasicrystal.
 (i) Same as (g) but for an Ammann-Beenker quasicrystal.
 A common colored-scale bar for (f-i) is shown at the right side of (i).
}
\end{figure}

To clarify the effect of quasiperiodicity on $|\braket{\hat{b}_i}|$ near a phase transition point, we compare a critical behavior of $|\braket{\hat{b}_i}|$ in crystals and quasicrystals for $\mu/U=0.5$.
First, we consider a square lattice.
In Fig.~\ref{fig:comparisonBeta}(a), we show $|\braket{\hat{b}_i}|^2$ as a function of $\overline{Z_i}J/U$ by a red line.
Since this red line is almost straight for larger $J$, we also plot a linear fitting by a green-dashed line.
By extrapolating this green-dashed line to $|\braket{\hat{b}_i}|=0$, we obtain a critical hopping strength $\overline{Z_i}J_{\rm c}/U \sim 0.1666$ for square lattice, as indicated by a downward-black arrow.
A critical exponent for $|\braket{\hat{b}_i}|$ is $1/2$~\cite{Fisher_1989,Sanders_2017} and we have
\begin{equation}
    |\braket{\hat{b}_i}|=a\sqrt{\frac{J-J_{\rm c}}{J_{\rm c}}},
    \label{eq:criticalBehavior}
\end{equation}
where the constant $a$ is approximately $1.31$ for square lattice.
Note that, in square lattice, all vertices share the same $|\braket{\hat{b}_i}|$ and thus a single $a$ is used for the entire system.

Next, we consider quasicrystals.
Figure~\ref{fig:comparisonBeta}(b) shows $|\braket{\hat{b}_i}|^2$ for various vertices in a Penrose quasicrystal, where we select one vertex located close to the center of system in the real system for each group shown in Fig.~\ref{fig:MISFboundarySi_Pen}(a).
As $J$ increases, $|\braket{\hat{b}_i}|^2$ becomes finite for all vertices simultaneously at $\overline{Z_i}J_{\rm c}/U \sim 0.1574$, and then linearly increases.
The slopes of these linear increases depend on types of vertices.
Figure~\ref{fig:comparisonBeta}(c) shows a result for an Ammann-Beenker quasicrystal, where we obtain $\overline{Z_i}J_{\rm c}/U \sim 0.1579$.
We also obtain a critical behavior of $|\braket{\hat{b}_i}|$ following Eq.~(\ref{eq:criticalBehavior}) regardless of a coordination number $Z_i$.

Figure~\ref{fig:comparisonBeta}(d) shows a relationship between $a$ and $Z_i$ in a Penrose quasicrystal.
The blue circles represent mean values of $a$ with respect to vertices with the same $Z_i$.
As these mean values are almost on a straight line, we show a linear fitting of them by a blue-dashed line.
It implies that $|\braket{\hat{b}_i}|$ near the phase transition point is almost proportional to the number of linked vertices $Z_i$.
This is reasonable because the mean field should be larger for a vertex with larger $Z_i$ [see Eq.~(\ref{eq:selfConsistent})].
A value of $a$ for a square lattice is indicated by a green-dashed line, which intersects with the blue-dashed line between $Z_i=4$ and 5.
Although the average value of $Z_i$ in Penrose quasicrystals is 4 which is the same as in square lattice, we find that $a$ for a square lattice is larger than $a$ for $Z_i=4$ in a Penrose quasicrystal.
We show a result for Ammann-Beenker quasicrystal in Fig.~\ref{fig:comparisonBeta}(e), where we also find a linear dependence of $Z_i$ on $a$.
The intersection of the blue-dashed and green-dashed lines is between $Z_i=4$ and 5 in an Ammann-Beenker quasicrystal.

To clarify the effect of quasiperiodicity on $a$, we show the real-space distribution of $a$ in a Penrose quasicrystal in Fig.~\ref{fig:comparisonBeta}(f).
A colored-scale bar is embedded at the far right, where a rightward-black arrow indicates $a$ for square lattice.
We find an intricate pattern of $a$ in the real space, however, a well-organized structure appears in the perpendicular space representation [Fig.~\ref{fig:comparisonBeta}(g)], where a blue-star (red-star) shaped region in $\tilde{z}=0$ ($1/\sqrt{2}$) is seen.
Within each domain of the same type of vertices, we find a non-uniform and ordered pattern.
For example, we find a yellow-triangular shaped region for D vertices (see the inset in Fig.~\ref{fig:MISFboundarySi_Pen}).
Accordingly, we confirm that quasiperiodicity governs the critical behavior of $|\braket{\hat{b}_i}|$ beyond the nearest-neighbor links.
Figures~\ref{fig:comparisonBeta}(h) and \ref{fig:comparisonBeta}(i) show the real-space and perpendicular-space distribution of $a$ in an Ammann-Beenker quasicrystal, respectively.
We also find that quasiperiodicity affects the critical behavior.

According to the above discussion, $|\braket{\hat{b}_i}|$ becomes finite for all vertices at the same hopping amplitude $J=J_{\rm c}$.
For a certain $(J_{\rm c}<)J$, $|\braket{\hat{b}_i}|$ for some vertices are larger than others because $a$ is different among vertices [see Figs.~\ref{fig:comparisonBeta}(b) and \ref{fig:comparisonBeta}(c)].
This explains that $J^{\rm c}_i$ in Figs.~\ref{fig:MISFboundarySi_Pen}(a) and \ref{fig:MISFboundarySi}(a) varies among vertices.
Also note that $J^{*}$ shown in Figs.~\ref{fig:criticalJ_N_combined_Pen}(a) and \ref{fig:criticalJ_N_combined}(a) becomes $J_{\rm c}$ for $\gamma_n\rightarrow 0$ as the critical hopping amplitude is shared by all vertices.
The origin of a large order metric at $J\sim J_{\rm c}$ in quasicrystals is attributed to a wide variety of $a$ and hence quasiperiodicity [see an inset of Fig.~\ref{fig:orderMetric}(f)].

\subsubsection{Comparison between amorphous systems and quasicrystals}

Our discussion regarding the hyperuniform analysis on a bosonic system can be applied to other systems that are neither crystals nor quasicrystals, namely systems without Bragg peaks.
We refer to such systems as amorphous systems.
We discuss below how the hyperuniform analysis contributes to differentiating between amorphous and quasicrystalline systems.

Some amorphous systems may not show the Mott insulating phase.
In fact, Halu {\it{et al}}.~\cite{Halu_2012} have shown that the Mott insulating phase is not present in random scale-free network systems in the thermodynamic limit.
Such systems are of course different from quasicrystalline systems, and we do not need to use the hyperuniform analysis.
Below, we consider systems that show the Mott insulating phase.

If the point distribution of such an amorphous system is not hyperuniform, hyperuniformity of a physical quantity can differentiate between amorphous and quasicrystals.
As we have discussed in Sec.~\ref{subsec:Hyp}, the order metric of a physical quantity approaches a value for the point distribution when $J$ increases.
However, we cannot define the order metric in amorphous because of its non-hyperuniform point distribution.
Therefore, the system should be multifractal.
This is in stark contrast to quasicrystals where the system should be hyperuniform when $J$ is large.

If the point distribution is hyperuniform, hyperuniformity of a physical quantity cannot differentiate between amorphous and quasicrystals.
We note that there exist hyperuniform amorphous systems~\cite{Zheng_2020, Morse_2023_Sep}.
Such amorphous systems should show a significant increase in the order metric near the phase boundary because each vertex is located on an environment different from other vertices, which is the same as in quasicrystals.
Even when it is not at the transition point, a physical quantity is inhomogeneous and the order metric is larger than that of the point distribution.
Because both amorphous and quasicrystalline systems share these properties, differentiating between them is difficult.

\section{With Disorder}
\label{sec:withDis}
In this section, we consider a random on-site potential with the uniform distribution $\epsilon_i \in [-\frac{\Delta}{2},\frac{\Delta}{2}]$, where $\Delta$ represents a strength of disorder.
With random on-site potentials, a bosonic system shows the Bose glass phase for $J\sim J_{\rm c}$ both in crystals and quasicrystals~\cite{Johnstone2021}.
To clarify how randomness affects the distribution of physical quantities, we apply multifractal analyses.
We calculate physical quantities for 30 shots of randomness, and we use the mean values of them. 

\subsection{Penrose quasicrystals}

\begin{figure}
	\includegraphics[width=8.6cm]{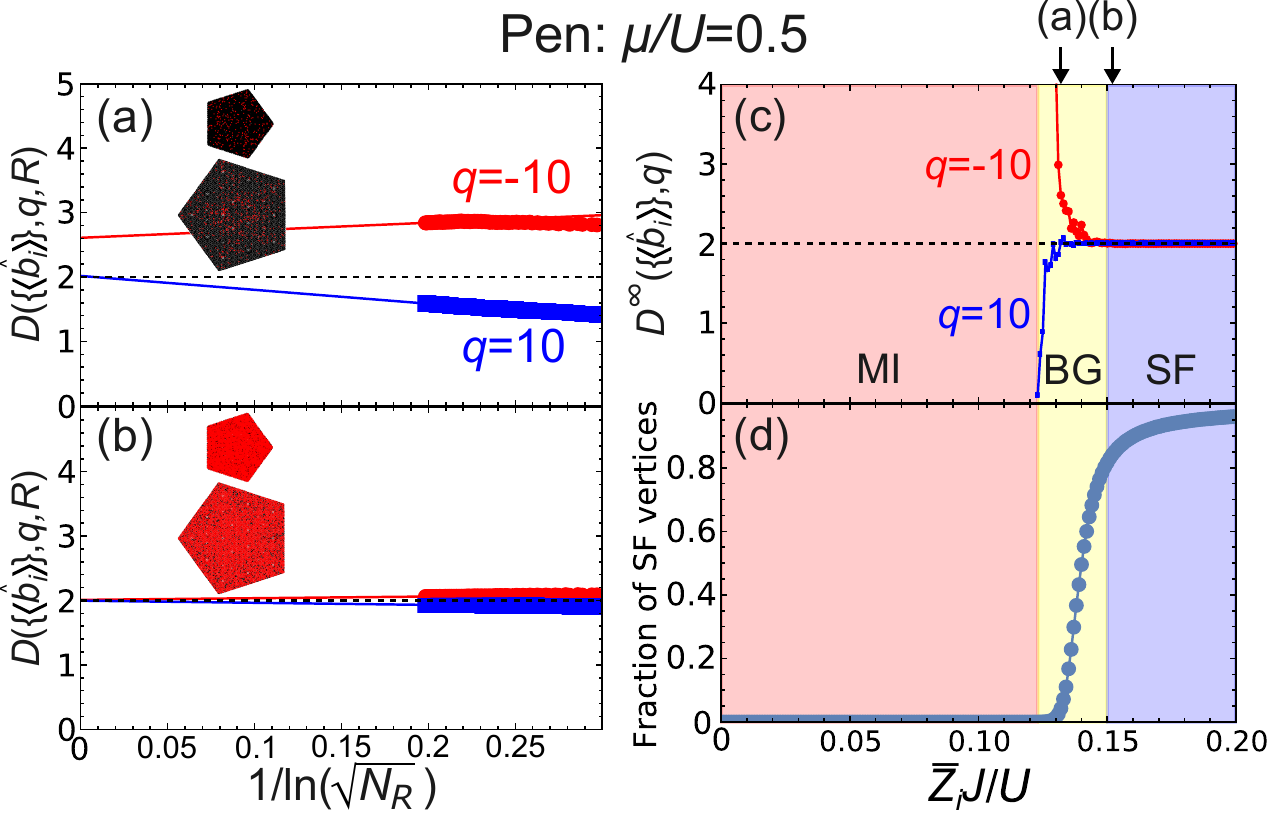}
	\caption{\label{fig:multiF_largeRandomness_AB_muIndex1_Pen_muIndex1}
 (Color online)
(a) The $R$-dependent multifractal dimension $D(\{ \braket{\hat{b}_i} \},q=-10,R)$ and $D(\{\braket{\hat{b}_i} \},q=10,R)$ as a function of $1/\ln(\sqrt{N_R})$ for $\Delta/U=0.6$, and $\overline{Z_i} J/U=0.132$ in a Penrose quasicrystal with $\mu/U=0.5$.
(b) Same as (a) but for $\overline{Z_i} J/U=0.151$.
(c) The $\overline{Z_i}J/U$ dependence of $D^{\infty}(\{ \braket{\hat{b}_i} \},q=-10)$ and $D^{\infty}(\{ \braket{\hat{b}_i} \},q=10)$.
MI, BG, and SF represent Mott insulating, Bose glass, and superfluid phases, respectively.
(d) The fraction of superfluid vertices as a function of $\overline{Z_i}J/U$.
We use 30 shots of randomness and average them, except for the insets in (a) and (b) where we show results for a single shot of randomness.
}
\end{figure}

First, we focus on Penrose quasicrystals.
Figure~\ref{fig:multiF_largeRandomness_AB_muIndex1_Pen_muIndex1}(a) shows $D(\{\braket{\hat{b}_i} \}, q, R)$ in Eq.~(\ref{eq:RDepMFD}) as a function of $1/\ln(\sqrt{N_R})$ for $q=-10$ (red circles) and $10$ (blue squares).
The parameter set used is $\mu/U=0.5, \Delta/U=0.6$, and $\overline{Z_i} J/U=0.132$.
By linearly fitting data points in Fig.~\ref{fig:multiF_largeRandomness_AB_muIndex1_Pen_muIndex1}(a), we obtain extrapolated values of $D(\{\braket{\hat{b}_i} \}, q, R)$ in the limit of $N_{R}\rightarrow \infty$.
The extrapolated $D^{\infty}(\{ \braket{\hat{b}_i} \},q=-10)$ and $D^{\infty}(\{ \braket{\hat{b}_i} \},q=10)$ deviate from 2.0, which indicates that this system is multifractal.
The inset of Fig.~\ref{fig:multiF_largeRandomness_AB_muIndex1_Pen_muIndex1}(a) shows the distribution of $S_i$ [see Eq.~(\ref{eq:SiDef})] in the perpendicular space for $\tilde{z}=0$ (upper panel) and $1/\sqrt{2}$ (lower panel).
We find that the distribution of superfluid vertices ($S_i=1$ represented by red dots) reflects quasiperiodic structure as a star-shaped region is relatively red in color, especially for $\tilde{z}=1/\sqrt{2}$.
However, this distribution does not completely obey quasiperiodic structure because of the presence of on-site random potentials.
As a result, there are randomly distributed Mott insulating vertices in the star-shaped region, whose distribution less reflects quasiperiodic structure for larger $\Delta/U$.
We refer to this phase as a Bose glass phase, not a superfluid phase.

Figure~\ref{fig:multiF_largeRandomness_AB_muIndex1_Pen_muIndex1}(b) is the same as \ref{fig:multiF_largeRandomness_AB_muIndex1_Pen_muIndex1}(a) but for $\overline{Z_i}J/U=0.151$.
Both $D^{\infty}(\{ \braket{\hat{b}_i} \},q=-10)$ and $D^{\infty}(\{ \braket{\hat{b}_i} \},q=10)$ are almost 2.0, which indicates a nonmultifractal behavior.
The inset clearly shows that this phase is a superfluid phase.
In Sec.~\ref{sec:ep0}, we have shown that the superfluid phase is hyperuniform, which agrees with Fig.~\ref{fig:multiF_largeRandomness_AB_muIndex1_Pen_muIndex1}(b).

From the above discussion, the Bose glass phase is multifractal while the superfluid phase is hyperuniform.
Also, the Mott insulating phase is hyperuniform as we have $\braket{\hat{b}_i}=0$ for all vertices.
To examine phase transition between these phases, we show the hopping amplitude dependence of $D^{\infty}(\{ \braket{\hat{b}_i} \},q=-10)$ (red circles) and $D^{\infty}(\{ \braket{\hat{b}_i} \},q=10)$ (blue squares) in Fig.~\ref{fig:multiF_largeRandomness_AB_muIndex1_Pen_muIndex1}(c) for the same parameter set as Figs.~\ref{fig:multiF_largeRandomness_AB_muIndex1_Pen_muIndex1}(a) and \ref{fig:multiF_largeRandomness_AB_muIndex1_Pen_muIndex1}(b).
The two downward-black arrows at the upper-right corner represent $\overline{Z_i}J/U$ used for Figs.~\ref{fig:multiF_largeRandomness_AB_muIndex1_Pen_muIndex1}(a) and \ref{fig:multiF_largeRandomness_AB_muIndex1_Pen_muIndex1}(b).
We do not show data for $\overline{Z_i}J/U \lesssim 0.125$ because $|\braket{\hat{b}_i}|$ is too small to calculate the multifractal dimension. 
Because both $D^{\infty}(\{ \braket{\hat{b}_i} \},q=-10)$ and $D^{\infty}(\{ \braket{\hat{b}_i} \},q=10)$ are away from 2.0, the system is multifractal for $0.125 \lesssim \overline{Z_i}J/U \lesssim $ 0.15.
On the other hand, the system is not multifractal for 0.15 $\lesssim \overline{Z_i}J/U$ as both $D^{\infty}(\{ \braket{\hat{b}_i} \},q=-10)$ and $D^{\infty}(\{ \braket{\hat{b}_i} \},q=10)$ are almost 2.0, and the system is hyperuniform according to Sec.~\ref{sec:ep0}.

We show the Mott insulating phase, the Bose glass phase, and the superfluid phase in red, yellow, and blue-shaded regions, respectively, in Fig.~\ref{fig:multiF_largeRandomness_AB_muIndex1_Pen_muIndex1}(c).
To determine phase boundaries, we consider the fraction of superfluid vertices as a function of $\overline{Z_i}J/U$ shown in Fig.~\ref{fig:multiF_largeRandomness_AB_muIndex1_Pen_muIndex1}(d).
For this calculation, we use 30 shots of randomness and average them.
The fraction is almost 1.0 for 0.15$ \lesssim \overline{Z_i}J/U$, where we assign a superfluid phase.
Comparing Figs.~\ref{fig:multiF_largeRandomness_AB_muIndex1_Pen_muIndex1}(c) and \ref{fig:multiF_largeRandomness_AB_muIndex1_Pen_muIndex1}(d), we conclude that our system changes from a hyperuniform Mott insulating phase to a multifractal Bose glass phase, and then to a hyperuniform superfluid phase with increasing $\overline{Z_i}J/U$.
We note that one can precisely determine the phase boundaries by, for example, a percolation analysis~\cite{Johnstone2021}.
The phase boundaries have been clarified for crystals in Refs.~\cite{Sengupta_2007_Aug,Bissbort_2009, Bissbort_2010, Niederle2013} and quasicrystals in Refs.~\cite{Gautier_2021,Johnstone2021,Ciardi_2023}.

\begin{figure}
	\includegraphics[width=8.6cm]{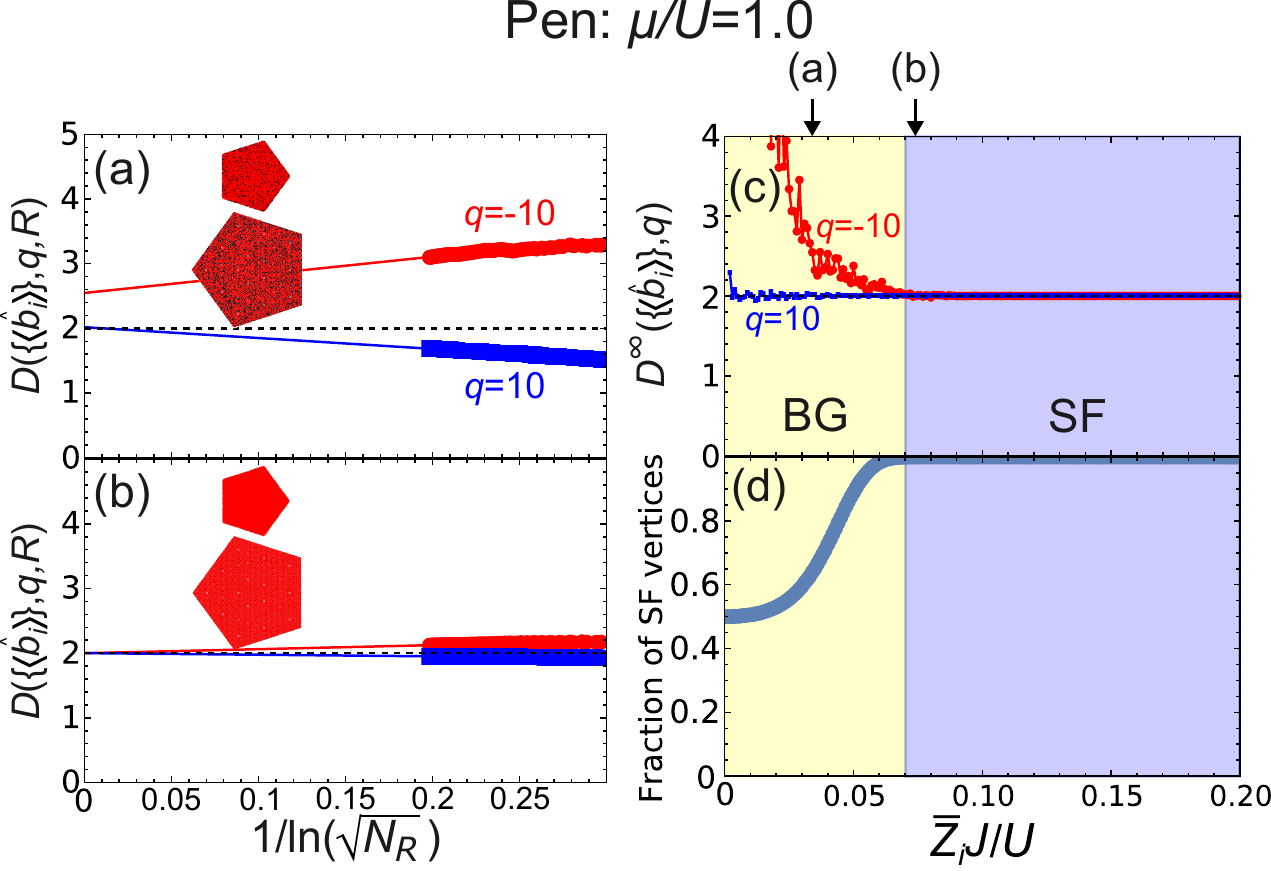}
	\caption{\label{fig:multiF_largeRandomness_AB_muIndex1_Pen_muIndex2}
 (Color online)
Same as Fig.~\ref{fig:multiF_largeRandomness_AB_muIndex1_Pen_muIndex1}(a) but for $\mu/U=1.0$. We use (a) $\overline{Z_i} J/U=0.034$ and (b) $0.074$.
}
\end{figure}

Figure~\ref{fig:multiF_largeRandomness_AB_muIndex1_Pen_muIndex2} shows results for $\mu/U=1.0$.
At $\overline{Z_i}J/U=0.034$, since both $D^{\infty}(\{ \braket{\hat{b}_i} \},q=-10)$ and $D^{\infty}(\{ \braket{\hat{b}_i} \},q=10)$ are away from 2.0, the system is multifractal as shown in Fig.~\ref{fig:multiF_largeRandomness_AB_muIndex1_Pen_muIndex2}(a).
The inset indicates that the system is a Bose glass phase.
At $\overline{Z_i}J/U=0.074$, both $D^{\infty}(\{ \braket{\hat{b}_i} \},q=-10)$ and $D^{\infty}(\{ \braket{\hat{b}_i} \},q=10)$ are almost 2.0, thus the system is not multifractal [see Fig.~\ref{fig:multiF_largeRandomness_AB_muIndex1_Pen_muIndex2}(b)].
As shown in the inset, most vertices are superfluid vertices and thus the system is a superfluid phase.
Combining the results of $D^{\infty}(\{ \braket{\hat{b}_i} \},q)$ in  Fig.~\ref{fig:multiF_largeRandomness_AB_muIndex1_Pen_muIndex2}(c) and the fraction of superfluid vertices in Fig.~\ref{fig:multiF_largeRandomness_AB_muIndex1_Pen_muIndex2}(d), we approximately assign that the system is (not) multifractal for $\overline{Z_i}J/U \lesssim 0.07$ ($0.07 \lesssim \overline{Z_i}J/U$).
For $\mu/U=1.0$, we find a multifractal Bose glass phase and a hyperuniform superfluid phase, but a Mott insulating phase does not appear.

\begin{figure}
	\includegraphics[width=8.6cm]{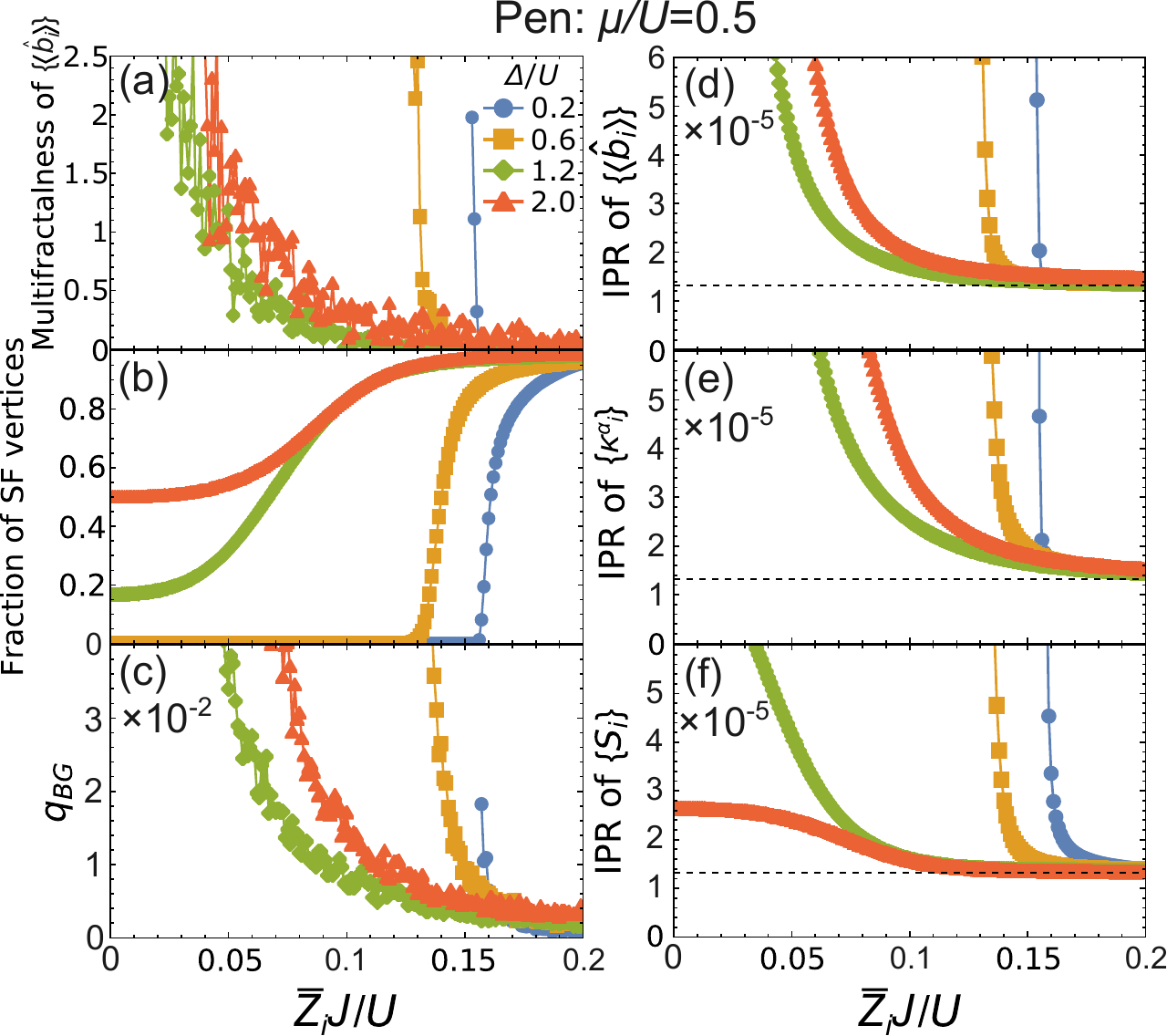}
	\caption{\label{fig:multifractalness_IPR_glassOrder_Pen_muIndex1}
 (Color online)
The hopping amplitude dependence of (a) the multifractalness of $\{ \braket{\hat{b}_i} \}$, (b) the fraction of superfluid vertices, (c) the replica overlap for Bose glass, (d) IPR of $\{\braket{\hat{b}_i}\}$, (e)  IPR of $\{\kappa_i^{\alpha}\}$, and (f) IPR of $\{S_i\}$ for $\mu/U=0.5$.
For these calculations, we use 30 shots of randomness and average them.
We consider all vertices $N=75806$ including an edge for (d-f), and the black-dashed lines indicate ${\rm IPR}=1/N$.
}
\end{figure}

Figure~\ref{fig:multifractalness_IPR_glassOrder_Pen_muIndex1}(a) shows multifractalness in $\{ \braket{\hat{b}_i} \}$, namely $D^{\infty}(\{ \braket{\hat{b}_i} \},q=-10)-D^{\infty}(\{ \braket{\hat{b}_i} \},q=10)$, as a function of $\overline{Z_i}J/U$, for $\mu/U=0.5$ and $\Delta/U=0.2$ (blue circles), 0.6 (orange squares), 1.2 (green diamonds), and 2.0 (red triangles).
We find that multifractalness decreases with increasing $\overline{Z_i}J/U$ regardless of $\Delta/U$.
Note that, when multifractalness is larger than 0, the system is multifractal.
We confirm that the system is hyperuniform for larger $\overline{Z_i}J/U$ irrespective of $\Delta/U$.

Figure~\ref{fig:multifractalness_IPR_glassOrder_Pen_muIndex1}(b) shows the fraction of superfluid vertices.
The minimum hopping amplitude for nonzero fraction at $\Delta/U=0.2$ and 0.6 is a critical hopping amplitude at which a phase transition between Mott insulating and Bose glass phases occurs.
The fraction increases as the hopping amplitude increases.
For $\Delta/U=1.2$ and 2.0, the fraction at $J=0$ is larger than 0 and thus the system is a Bose glass phase, which implies that a Mott insulating phase does not appear for large $\Delta/U$.
Then we have only one phase transition from a multifractal Bose glass phase to a hyperuniform superfluid phase.
Since multifractalness is larger at $\Delta/U=2.0$ than 1.2 as shown in Fig.~\ref{fig:multifractalness_IPR_glassOrder_Pen_muIndex1}(a), a multifractal phase, i.e., a Bose glass phase, becomes stable for large $\Delta$, leading to disappearance of a Mott insulating phase.

Figure~\ref{fig:multifractalness_IPR_glassOrder_Pen_muIndex1}(c) shows a replica overlap for Bose glass $q_\mathrm{BG}$ in Eq.~(\ref{eq:qBG13}).
$q_\mathrm{BG}$ is nonzero in Bose glass phase, but almost zero in superfluid phase.
This indicates that, in Bose glass phase, the fluctuation of $\braket{\hat{n}_i}$ overlaps significantly among different kinds of randomness.

Figures~\ref{fig:multifractalness_IPR_glassOrder_Pen_muIndex1}(d), \ref{fig:multifractalness_IPR_glassOrder_Pen_muIndex1}(e), and \ref{fig:multifractalness_IPR_glassOrder_Pen_muIndex1}(f) show IPR of $\{\braket{\hat{b}_i}\}$, $\{\kappa_i^{\alpha}\}$, and $\{S_i\}$, respectively.
The black-dashed lines indicate ${\rm IPR}=1/N$, which is the allowed minimum value of IPR.
For $J\rightarrow \infty$, all the vertices share almost the same value of physical quantities, which makes IPR the minimum.
The $J$ and $\Delta$ dependencies are similar to those in Fig.~\ref{fig:multifractalness_IPR_glassOrder_Pen_muIndex1}(a), except for $\{S_i\}$.
Since IPR is a measure of localization for physical quantities, the enhancement of multifractalness is closely related to a localized nature of the system.

\begin{figure}
	\includegraphics[width=8.6cm]{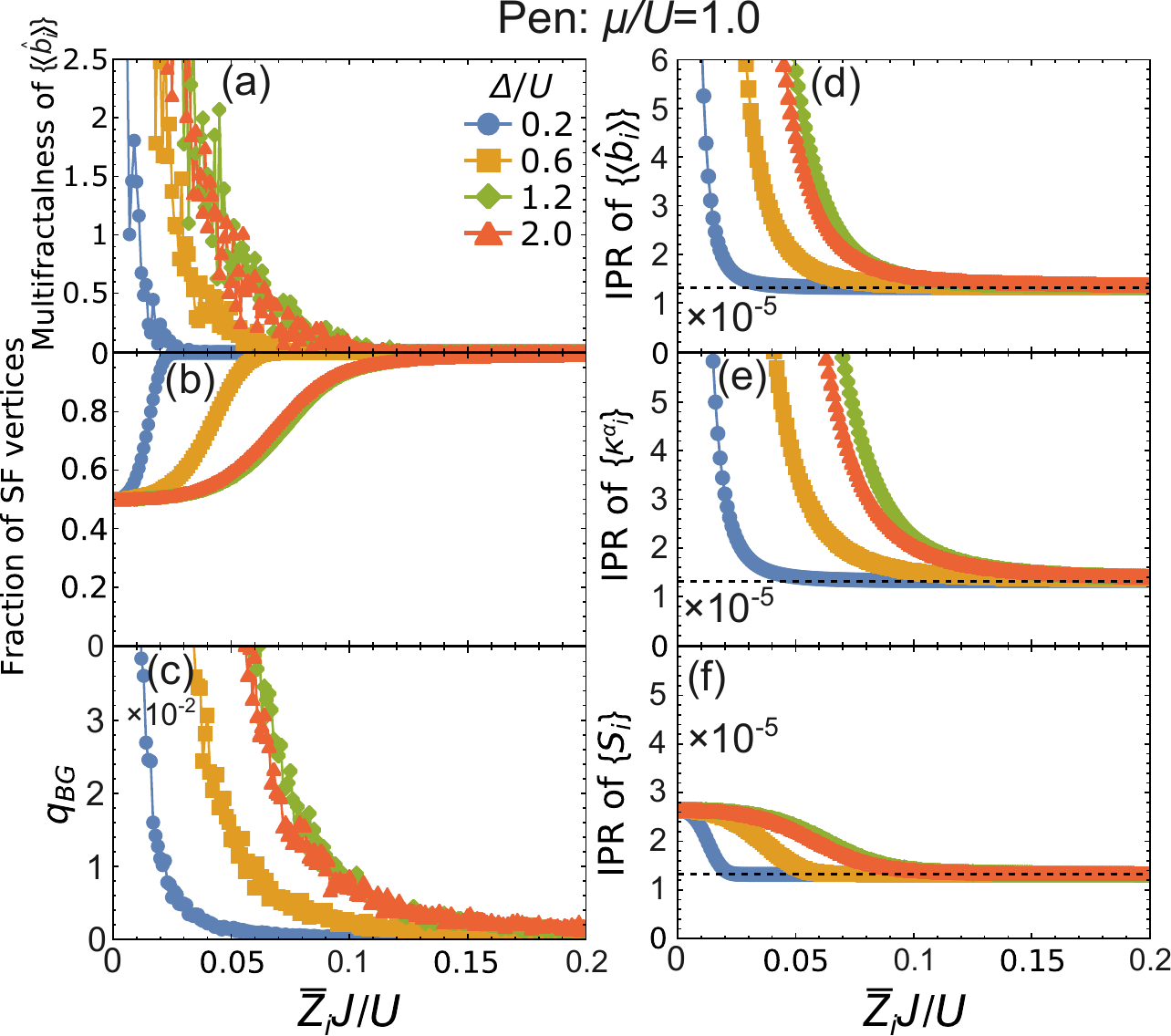}
	\caption{\label{fig:multifractalness_IPR_glassOrder_Pen_muIndex2}
 (Color online)
Same as Fig.~\ref{fig:multifractalness_IPR_glassOrder_Pen_muIndex1} but for $\mu/U=1.0$.
}
\end{figure}

Figure~\ref{fig:multifractalness_IPR_glassOrder_Pen_muIndex2} is the same as Fig.~\ref{fig:multifractalness_IPR_glassOrder_Pen_muIndex1} but for $\mu/U=1.0$.
We find that multifractalness in Fig.~\ref{fig:multifractalness_IPR_glassOrder_Pen_muIndex2}(a) decreases with increasing $\overline{Z_i}J/U$ and the system is multifractal (hyperuniform) for small (large) $\overline{Z_i}J/U$ regardless of $\Delta/U$.
We do not find the Mott insulating phase in Fig.~\ref{fig:multifractalness_IPR_glassOrder_Pen_muIndex2}(b) as the fraction of superfluid vertices is larger than zero.
Comparing Figs.~\ref{fig:multifractalness_IPR_glassOrder_Pen_muIndex2}(a) and \ref{fig:multifractalness_IPR_glassOrder_Pen_muIndex2}(b), we find that multifractalness is (not) larger than zero when the system is a Bose glass (superfluid) phase.
The behavior of replica overlap in Fig.~\ref{fig:multifractalness_IPR_glassOrder_Pen_muIndex2}(c) is similar to multifractalness in Fig.~\ref{fig:multifractalness_IPR_glassOrder_Pen_muIndex2}(a).
The replica overlap is large (small) in Bose glass phase (superfluid phase).
IPR of $\{\braket{\hat{b}_i}\}$, $\{\kappa_i^{\alpha}\}$, and $\{S_i\}$ in Figs.~\ref{fig:multifractalness_IPR_glassOrder_Pen_muIndex2}(d), \ref{fig:multifractalness_IPR_glassOrder_Pen_muIndex2}(e), and \ref{fig:multifractalness_IPR_glassOrder_Pen_muIndex2}(f) also show a $\Delta$ dependence similar to Fig.~\ref{fig:multifractalness_IPR_glassOrder_Pen_muIndex2}(a).
Similar to the $\mu/U=0.5$ case, physical quantities are localized when multifractalness is enhanced.

\subsection{Ammann-Beenker quasicrystals}

\begin{figure}
	\includegraphics[width=8.6cm]{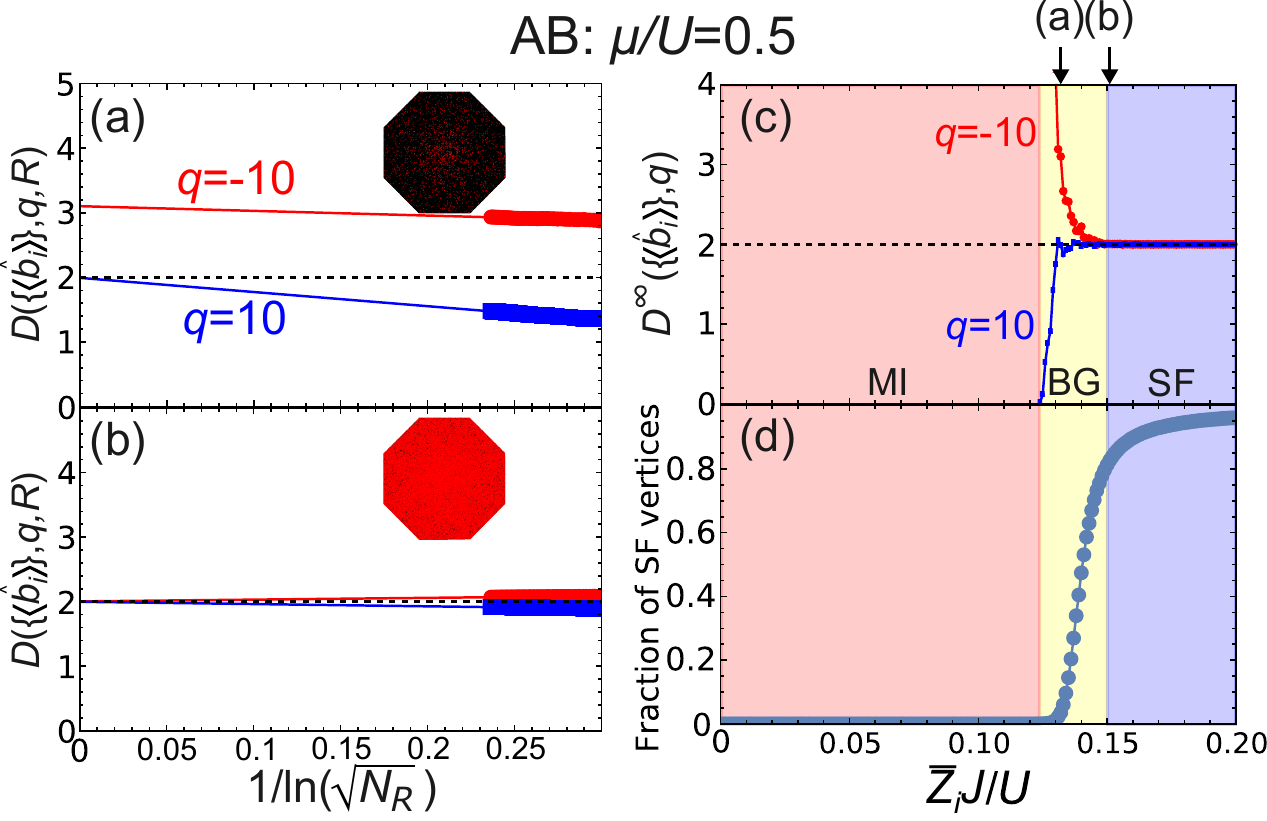}
	\caption{\label{fig:multiF_largeRandomness_AB_muIndex1}
 (Color online)
 Same as Fig.~\ref{fig:multiF_largeRandomness_AB_muIndex1_Pen_muIndex1} but for Ammann-Beenker quasicrystals with $\mu/U=0.5$.
 We use (a) $\overline{Z_i}J/U=0.132$ and (b) 0.151. 
}
\end{figure}

Next, we focus on Ammann-Beenker quasicrystals.
Figure~\ref{fig:multiF_largeRandomness_AB_muIndex1}(a) is the same as Fig.~\ref{fig:multiF_largeRandomness_AB_muIndex1_Pen_muIndex1}(a) but for Ammann-Beenker quasicrystals with $\mu/U=0.5$ and $\overline{Z_i} J/U=0.132$.
As both $D^{\infty}(\{ \braket{\hat{b}_i} \},q=-10)$ and $D^{\infty}(\{ \braket{\hat{b}_i} \},q=10)$ deviate from 2.0, the system is multifractal.
The inset where red dots distribute implies that the system is a Bose glass phase.
Figure~\ref{fig:multiF_largeRandomness_AB_muIndex1}(b) is for $\overline{Z_i} J/U=0.151$.
As both $D^{\infty}(\{ \braket{\hat{b}_i} \},q=-10)$ and $D^{\infty}(\{ \braket{\hat{b}_i} \},q=10)$ are almost 2.0, the system is not multifractal.
According to the inset where all vertices are superfluid vertex, the system is a hyperuniform superfluid phase.
From Figs.~\ref{fig:multiF_largeRandomness_AB_muIndex1}(c) and \ref{fig:multiF_largeRandomness_AB_muIndex1}(d), we find that the system changes from a hyperuniform Mott insulating phase to a multifractal Bose glass phase, and then to a hyperuniform superfluid phase with increasing $\overline{Z_i}J/U$.

\begin{figure}
	\includegraphics[width=8.6cm]{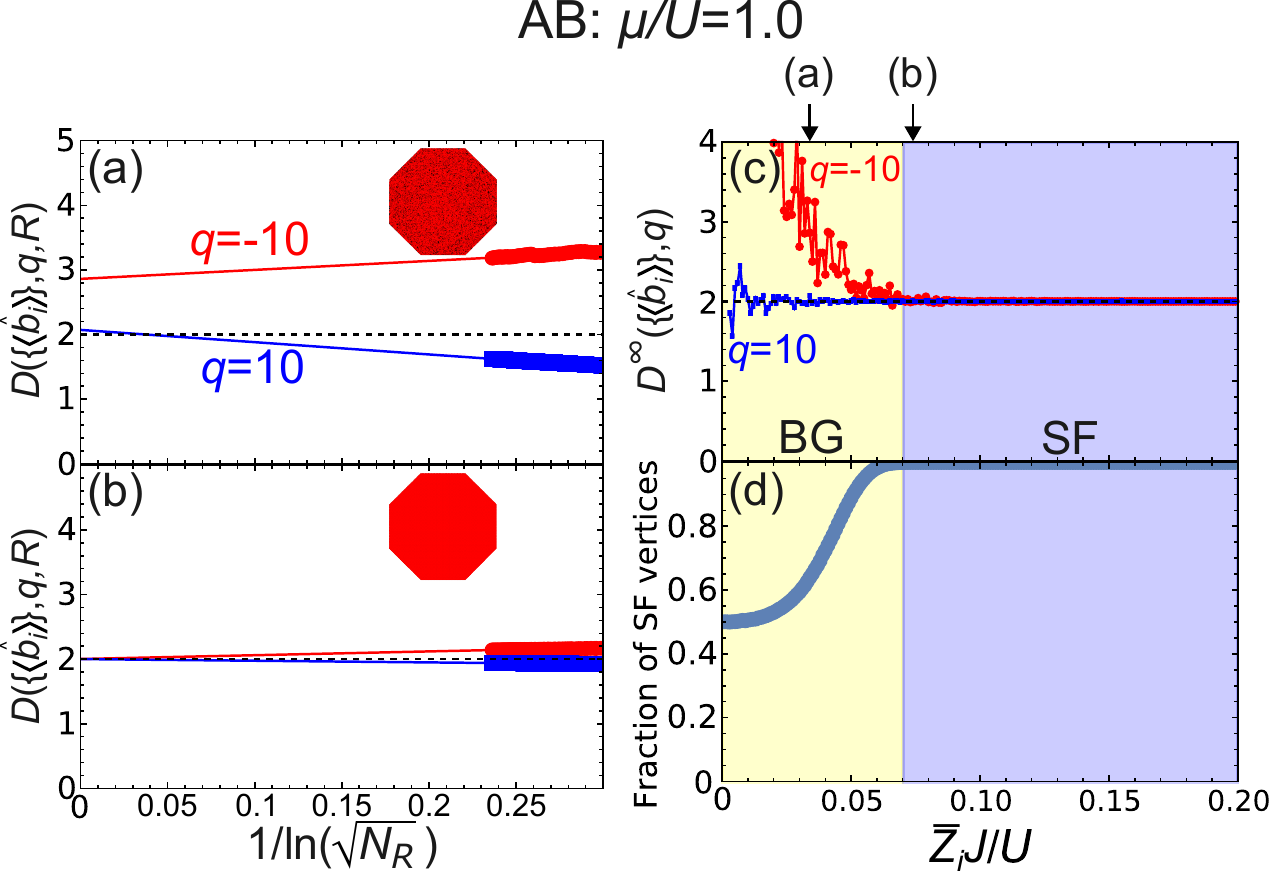}
	\caption{\label{fig:multiF_largeRandomness_AB_muIndex2}
 (Color online)
 Same as Fig.~\ref{fig:multiF_largeRandomness_AB_muIndex1} but for $\mu/U=1.0$.
 We use (a) $\overline{Z_i}J/U=0.034$ and (b) 0.074. 
}
\end{figure}

Figure~\ref{fig:multiF_largeRandomness_AB_muIndex2} is the same as Fig.~\ref{fig:multiF_largeRandomness_AB_muIndex1} but for $\mu/U=1.0$.
We find that the system is a multifractal Bose glass (hyperuniform superfluid) phase for $\overline{Z_i}J/U=0.034$ and 0.074 from Figs.~\ref{fig:multiF_largeRandomness_AB_muIndex2}(a) and \ref{fig:multiF_largeRandomness_AB_muIndex2}(b).
Figures~\ref{fig:multiF_largeRandomness_AB_muIndex2}(c) and \ref{fig:multiF_largeRandomness_AB_muIndex2}(d) show that the system changes from a multifractal Bose glass phase to a hyperuniform superfluid phase with increasing $\overline{Z_i}J/U$.

\begin{figure}
	\includegraphics[width=8.6cm]{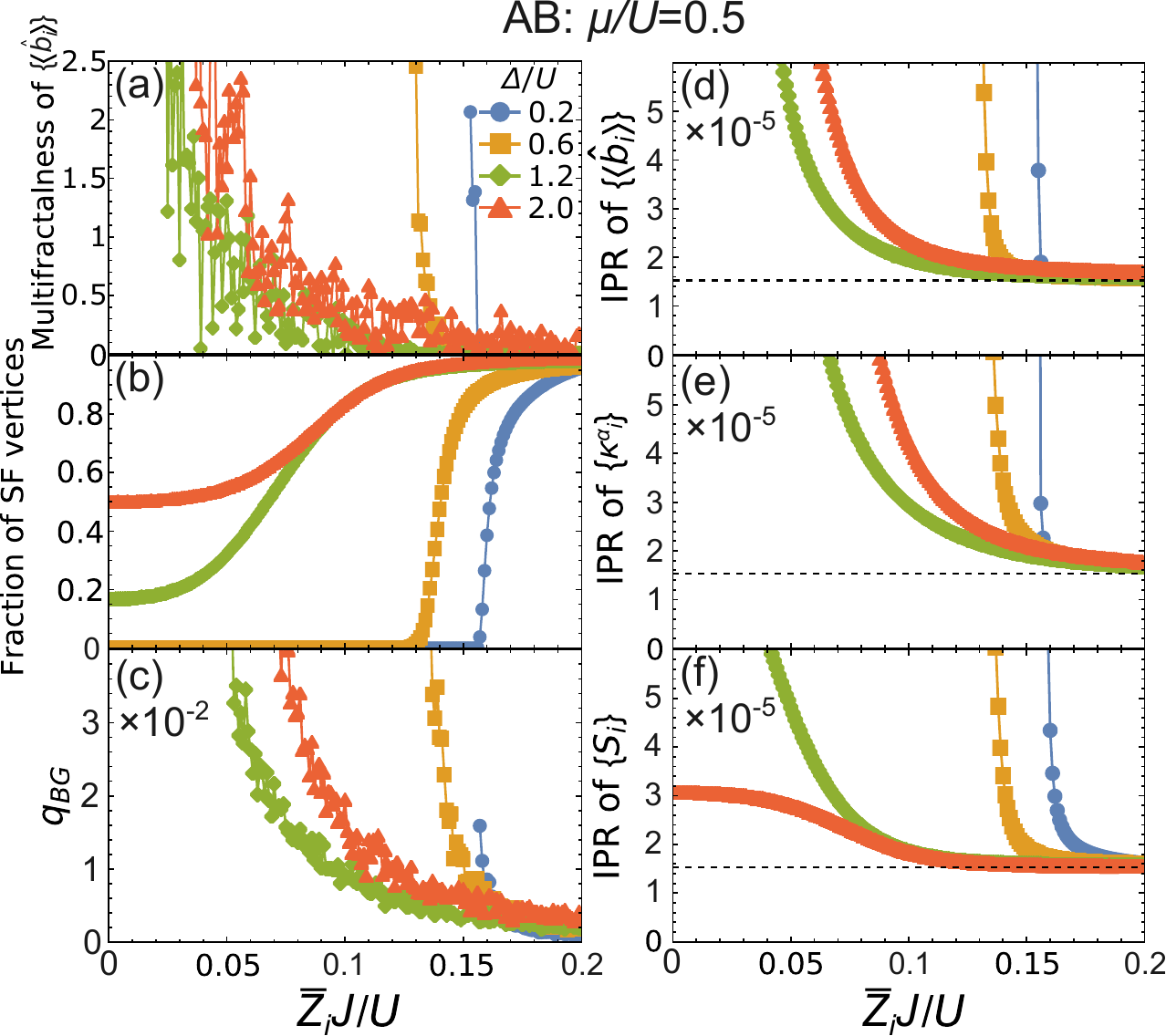}
	\caption{\label{fig:multifractalness_IPR_glassOrder}
 (Color online)
 Same as Fig.~\ref{fig:multifractalness_IPR_glassOrder_Pen_muIndex1} but for Ammann-Beenker quasicrystals with $\mu/U=0.5$.
 We consider all vertices $N=65281$ including an edge for (d-f), and the black-dashed lines indicate ${\rm IPR}=1/N$.
}
\end{figure}

\begin{figure}
	\includegraphics[width=8.6cm]{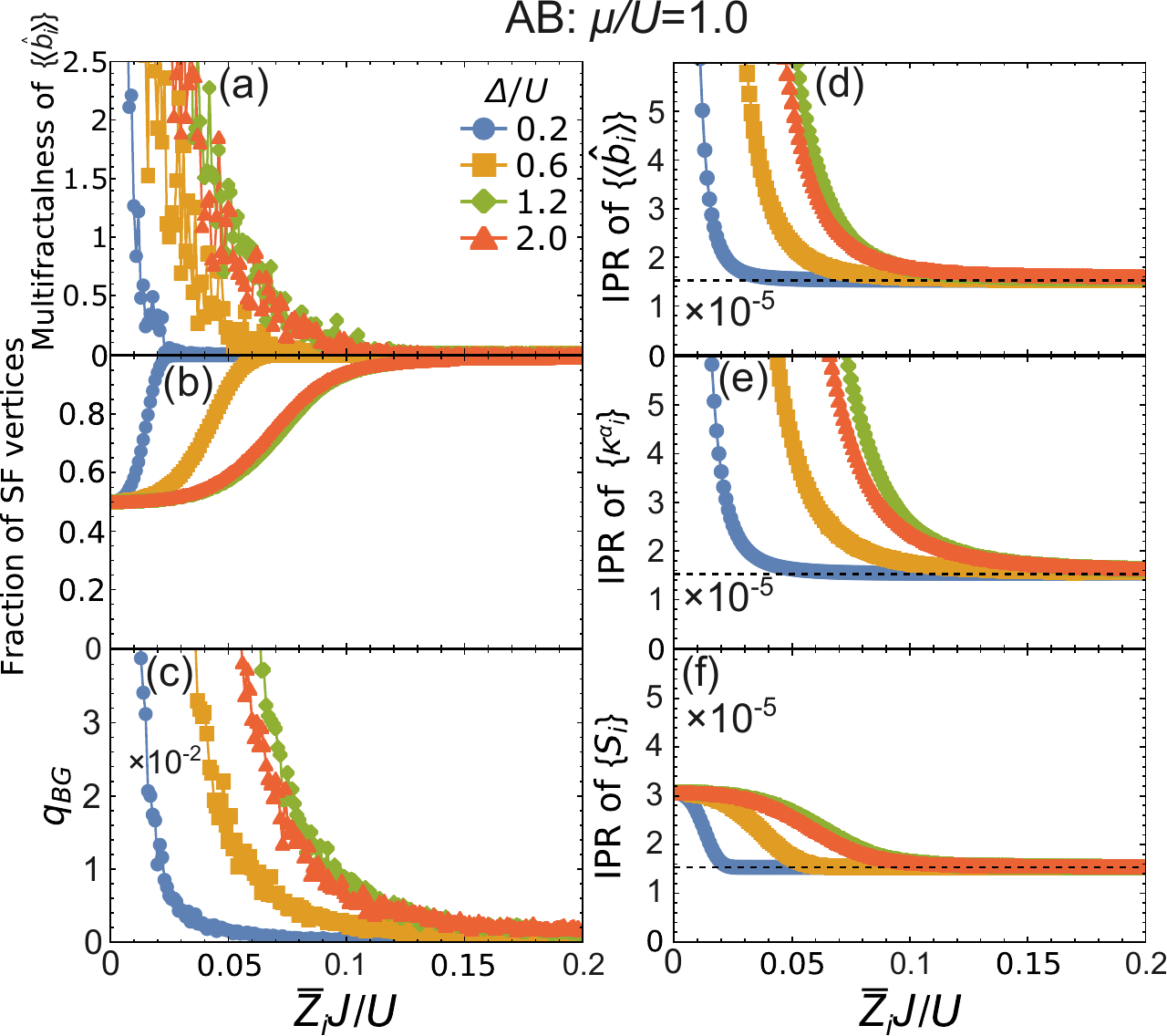}
	\caption{\label{fig:multifractalness_IPR_glassOrder_AB_muIndex2}
 (Color online)
 Same as Fig.~\ref{fig:multifractalness_IPR_glassOrder} but for $\mu/U=1.0$.
}
\end{figure}

Figures~\ref{fig:multifractalness_IPR_glassOrder} and \ref{fig:multifractalness_IPR_glassOrder_AB_muIndex2} are the same as Figs.~\ref{fig:multifractalness_IPR_glassOrder_Pen_muIndex1} and \ref{fig:multifractalness_IPR_glassOrder_Pen_muIndex2}, respectively,  but for Ammann-Beenker quasicrystals.
Both the $\mu/U=0.5$ and 1.0 exhibit almost the same behavior between Penrose and Ammann-Beenker quasicrystals in several quantities shown in the figures.
This is a striking observation that similarity between the two quasicrystals in various quantities is maintained even in the presence of disorder.

\subsection{Comparison between quasicrystals and crystals}

\begin{figure}
	\includegraphics[width=8.4cm]{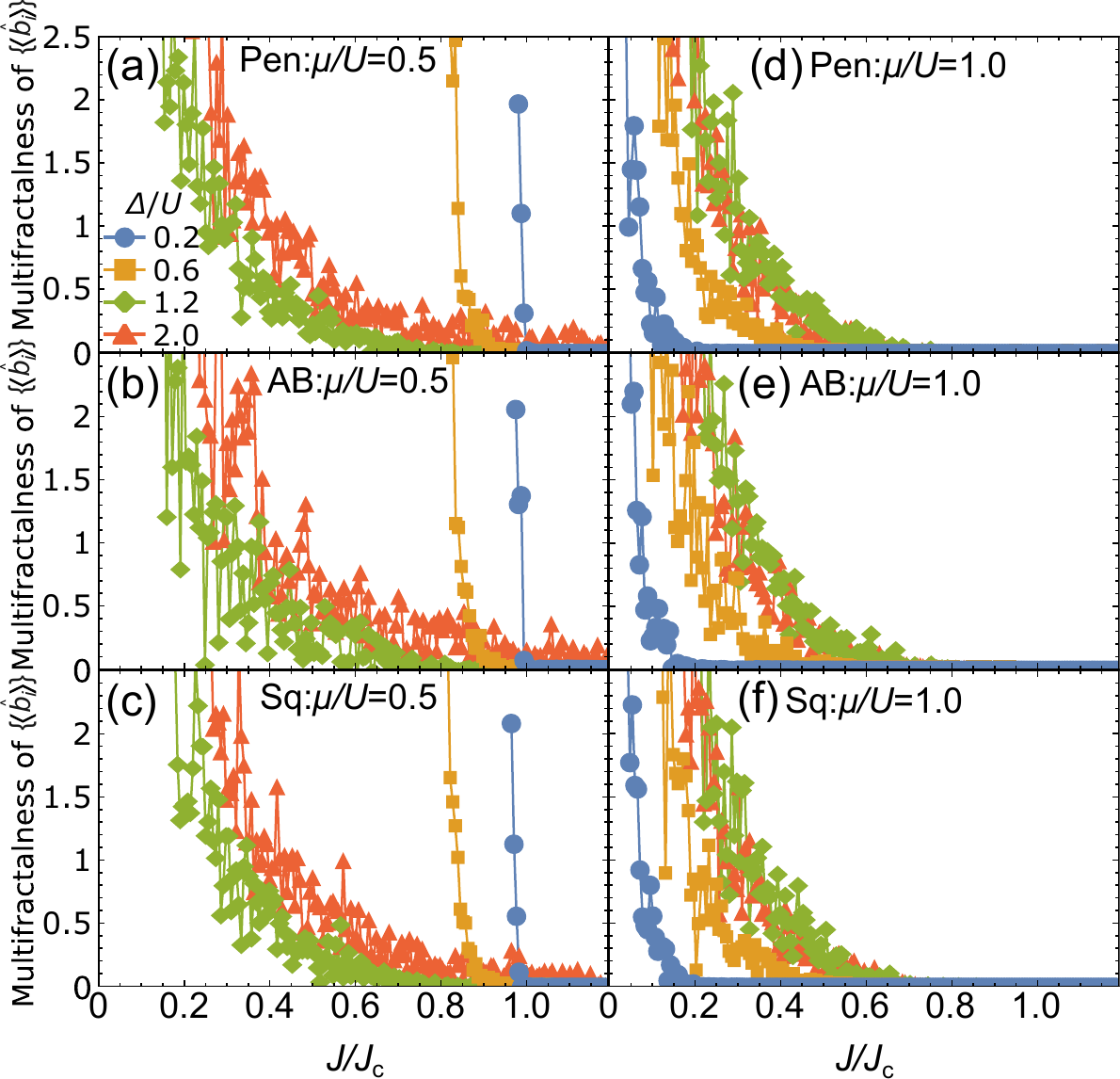}
	\caption{\label{fig:multifractalness_Pen_AB_Sq}
(Color online)
Scaled hopping amplitude dependence of multifractalness for various disorder $\Delta$. For $\mu/U=0.5$, (a) Penrose quasicrystals, (b) Ammann-Beenker quasicrystals, and (c) square lattice, and for $\mu/U=1.0$, (d) Penrose quasicrystals, (e) Ammann-Beenker quasicrystals, and (f) square lattice.
For these calculations, we use 30 shots of randomness and average them.
}
\end{figure}

To clarify the effect of quasiperiodicity on the multifractal property of a system, we compare multifractalness in quasicrystals and crystals.
Without disorder $\epsilon_i=0$, a critical hopping amplitude separating Mott insulating and superfluid phases, namely $J_{\rm c}$, is different between quasicrystals and crystals.
To make effect of this difference smaller, we normalize hopping amplitudes by $J_{\rm c}$ for both systems.
We show in Fig.~\ref{fig:multifractalness_Pen_AB_Sq} multifractalness of $\{ \braket{\hat{b}_i} \}$ as a function of $J/J_{\rm c}$ in the presence of disorder $\Delta$ for Penrose quasicrystals, Ammann-Beenker quasicrystals, and square lattice.
Multifractalness does not show any significant difference between quasicrystals and periodic lattice, for both $\mu/U=0.5$ and 1.0.

\begin{table*}[t]
\caption{
$N$, $N_{\rm in}$, and $R_{\rm in}/D$ for each generation in Penrose (Pen) and Ammann-Beenker (AB) quasicrystals.
}
\label{table:genPenAB}
\begin{tabular}{llllllllllll}
\hline\hline
& \multicolumn{1}{l|}{generation} & 1  & 2  & 3   & 4    & 5     & 6     & 7    & 8     & 9     & 10    \\
\hline
\CenterRow{3}{Pen} & 
\multicolumn{1}{l|}{$N$} & 11 & 31 & 86  & 226  & 601   & 1591  & 4181 & 11006 & 28901 & 75806 \\
\multicolumn{1}{c}{}                   & 
\multicolumn{1}{l|}{$N_{\rm in}$} &  & 16 & 51  & 141  & 401 & 861 & 2421 & 6171 & 18986 & 55651 \\
\multicolumn{1}{c}{}                   & 
\multicolumn{1}{l|}{$R_{\rm in}/D$} &  & 2 & 4  & 6  & 10 & 15 & 25 & 40 & 70 & 120 \\
\hline
\CenterRow{3}{AB} & 
\multicolumn{1}{l|}{$N$} &17 & 65 & 353 & 1969 & 11281 & 65281 \\
\multicolumn{1}{c}{}                   & 
\multicolumn{1}{l|}{$N_{\rm in}$} &  & 65 & 193  & 1185  & 6825 & 42649 \\
\multicolumn{1}{c}{}                   & 
\multicolumn{1}{l|}{$R_{\rm in}/D$} &  & 7$/\sqrt{2}$ & 10$/\sqrt{2}$  & 25$/\sqrt{2}$  & 60$/\sqrt{2}$ & 150$/\sqrt{2}$\\
\hline\hline
                                       &                          &    &    &     &      &       &       &      &       &       &      
\end{tabular}
\end{table*}

\section{Conclusion}
\label{sec:conclusion}

In summary, we have investigated multifractal or hyperuniform properties in the Bose-Hubbard model on the Penrose and Ammann-Beenker tilings with and without disorder.
We have found that, without disorder, the system is hyperuniform in both Mott insulating and superfluid phases.
From the enhancement of order metric, we have found that the complexity of the system increases near the phase boundary of the two phases.
This is in contrast to a constant order metric in square lattice.
A wider variety of vertices in quasicrystals allows more intricate distribution of a physical quantity, which results in the larger order metric than in crystals.
The difference between quasicrystals and crystals is more significant as approaching a phase boundary because of an increase in a typical correlation length of a physical quantity.
Examining a critical hopping strength for generating superfluid vertex near the phase boundary, we have found that the distribution of a physical quantity near a phase boundary reflects the quasiperiodicity of the point distribution in quasicrystals.
Examining a critical hopping strength for generating superfluid vertex near the phase boundary, we have attributed the origin of enhanced order metric to quasiperiodicity in the systems.
This contrasting behavior between crystalline and quasicrystalline systems can be confirmed by using ultracold atoms if one measures the distribution of local superfluid amplitude in optical lattice~\cite{Sbroscia_2020_Nov,yu_2022_thesis,Gottlob_2023_Apr, yu_2023}.

With disorder, we have found that the system changes from a hyperuniform Mott insulating phase to a multifractal Bose glass phase, and then to a hyperuniform superfluid phase as hopping amplitude increases.
We emphasize that, as far as we know, this is the first report for the phase transition between hyperuniform and multifractal systems.
We find no significant differences between the multifractal properties of the Bose glass phase in quasiperiodic and periodic systems.
Accordingly, quasiperiodicity is not essential for realizing multifractality in the presence of disorder.

In this study, underlying phase transitions are not quantitatively discussed because we have analyzed a fraction of superfluid vertices.
Future studies could explore this issue further by using, for example, a percolation analysis~\cite{Johnstone2021} and the cluster mean-field approximation~\cite{Gaude_2022_Jun}.
Also, future research on the phasons might extend the explanations of the effect of quasiperiodicity on multifractality and hyperuniformity~\cite{Rajagopal_2019_Nov}.
The phasons are a kind of randomness, related to the hyperspace of quasicrystals, and are realizable in quasicrystals only.
The Bose glass phase might be induced by the increases of the phasons in quasicrystals.

\section*{Acknowledgment}
\begin{acknowledgments}
This work was supported by the Natural Sciences and Engineering Research Council of Canada, JST SPRING (Grant No. JPMJSP2151) and the Japan Society for the Promotion of Science, KAKENHI (Grant No. JP19H05821).
\end{acknowledgments}

\appendix
\section{Inflation Rule}
\label{sec:AppDeflation}

\begin{figure}[t]
	\includegraphics[width=8.6cm]{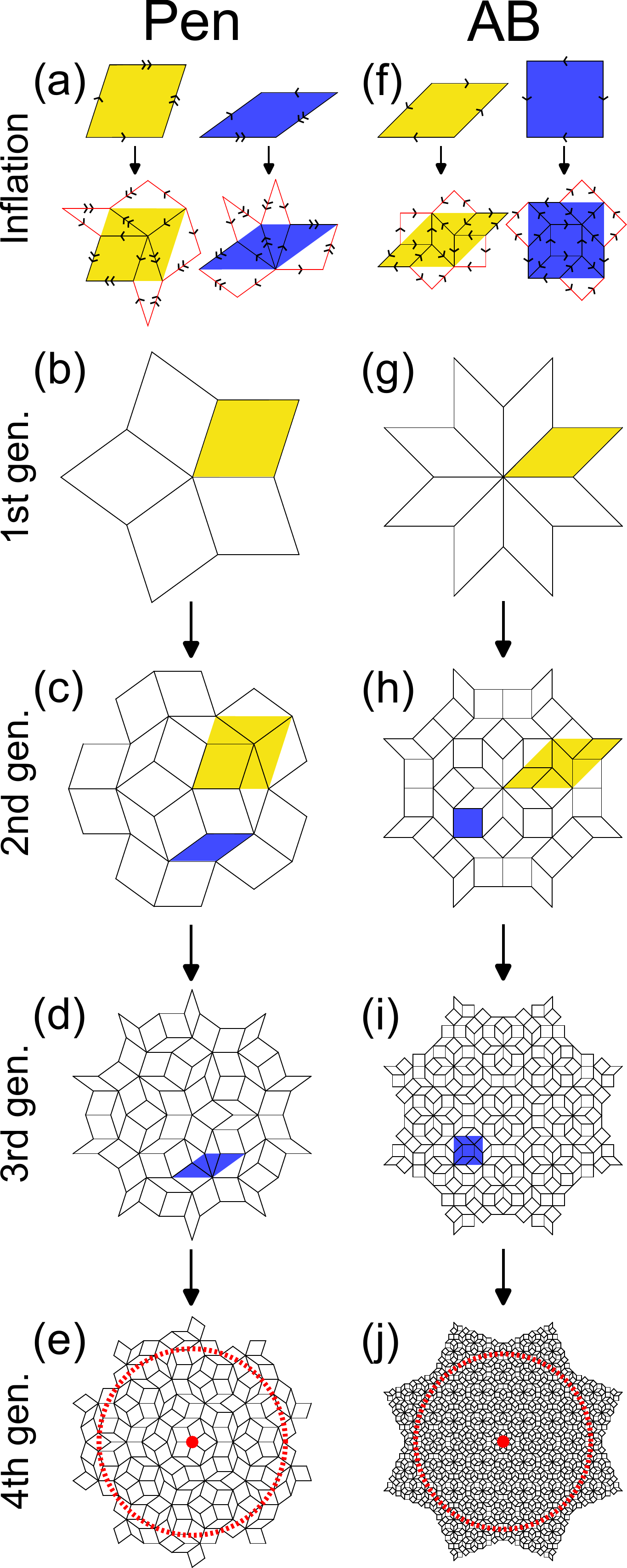}
	\caption{\label{fig:appPenAB}
 (Color online)
 Inflation rule (a) [(f)], tiling of 1st (b) [(g)], 2nd (c) [(h)], 3rd (d) [(i)], and 4th (e) [(j)] generations for Penrose [Ammann-Beenker] tilings.
 }
\end{figure}

We generate Penrose and Ammann-Beenker tilings according to inflation method.
An inflation for a Penrose tiling is an operation of substitution illustrated in Fig.~\ref{fig:appPenAB}(a).
A Penrose tiling contains two types of rhombi: a ``fat'' rhombus (left upper panel) and a ``skinny'' rhombus (right upper panel).
Note that each edge of these rhombi is distinguished by symbols of single and double arrows.
When an inflation operation acts on a tiling of $g$th generation, each component of the tiling in upper panels is replaced with a set of tiles in lower panels.
We call the resulting tiling a tiling of $(g+1)$th generation.
The red edges of the tiles in lower panels are generated by inflations applied to adjacent tiles.

We use a set of tiles in Fig.~\ref{fig:appPenAB}(b) as 1st generation of Penrose tiling.
By applying an inflation operation to Fig.~\ref{fig:appPenAB}(b), we obtain Fig.~\ref{fig:appPenAB}(c) as a tiling of 2nd generation.
A yellow fat rhombus in Fig.~\ref{fig:appPenAB}(b) results in a yellow-shaded region in Fig.~\ref{fig:appPenAB}(c).
Similarly, a blue skinny rhombus in Fig.~\ref{fig:appPenAB}(c) results in a blue-shaded region in a tiling of 3rd generation illustrated in Fig.~\ref{fig:appPenAB}(d).

We use open boundary condition.
One can remove effects from edges by considering the vertices inside a circular window near the center only.
For example, in Fig.~\ref{fig:appPenAB}(e), we show a red-dashed circular window with radius $R_{\rm in}/D=6$ whose center is indicated by a red small circle.
The number of vertices inside the circular window is $N_{\rm in}=141$.
Table~\ref{table:genPenAB} summarizes $N$,$N_{\rm in}$, and $R_{\rm in}/D$ for each generation, which we use for calculations in the main text.
Figures~\ref{fig:appPenAB}(f-j) are the same as Figs.~\ref{fig:appPenAB}(a-e) but for Ammann-Beenker quasicrystals.
We have $N=1969$, $N_{\rm in}=1185$, and $R_{\rm in}/D=25/\sqrt{2}$ in Fig.~\ref{fig:appPenAB}(j).



\end{document}